\documentclass[a4paper,11pt]{article}
\usepackage{pos}
\usepackage{subcaption}

\title{Paleo-Detectors as a Novel Probe of Dark Matter–Nucleus Effective Interactions}

\author*[a]{Dionysios P. Theodosopoulos}

\affiliation[a]{Weinberg Institute for Theoretical Physics, Department of Physics, The University of Texas at Austin,\\
  Austin, TX 78712, USA}

\emailAdd{d.theodosopoulos@utexas.edu}

\abstract{Paleo-detectors are a proposed approach to the direct detection of dark matter (DM) based on the observation of DM-induced nuclear recoils. Rather than relying on large detector masses required for conventional detectors, paleo-detectors exploit extremely long integration times, using small samples of naturally occurring minerals that have resided deep underground and accumulated damage tracks over geological timescales of order a gigayear. In this contribution, we review recent theoretical predictions for the sensitivity of paleo-detectors to WIMP–nucleus interactions within the framework of a Non-Relativistic Effective Field Theory (NREFT). We discuss both elastic and inelastic scattering processes and consider isoscalar couplings, taking into account backgrounds from neutrinos and radiogenic sources. We present projected sensitivities of paleo-detectors for different readout and exposure scenarios and compare the reach of paleo-detectors to that of conventional direct-detection experiments.
For DM masses in the range $1\, \mathrm{GeV}/c^2$--$10\, \mathrm{GeV}/c^2$, paleo-detectors are expected to achieve sensitivities exceeding those of conventional experiments for WIMP--nucleus interactions mediated by all NREFT operators, with only a weak dependence on the read-out scenario or target mineral. For larger DM masses, $10\, \mathrm{GeV}/c^2$--$5\, \mathrm{TeV}/c^2$, paleo-detectors are projected to attain sensitivities comparable to or surpassing conventional experiments for interactions mediated by several NREFT operators, depending on the specific read-out scenario and choice of target mineral.}

\FullConference{Proceedings of the Corfu Summer Institute 2025 "School and Workshops on Elementary Particle Physics and Gravity" (CORFU2025)\\
27 April - 28 September, 2025\\
Corfu, Greece\\}

\begin{document}
\maketitle

\section{Introduction}

A wide range of cosmological and astrophysical observations indicate that non-baryonic cold dark matter constitutes the dominant component of the matter content of the Universe, yet its fundamental nature remains unknown. Among the many proposed candidates, Weakly Interacting Massive Particles (WIMPs) are well motivated by a broad class of Beyond the Standard Model scenarios and are actively searched for in direct-detection experiments.

Paleo-detectors provide an alternative approach to direct dark matter detection by exploiting geological timescales to achieve extremely large effective exposures~\cite{Snowden-Ifft:1995zgn,Baum:2018tfw,Drukier:2018pdy,Edwards:2018hcf,Acevedo:2021tbl,Baum:2021jak,Fung:2025cub,Baltz:1997dw,Baum:2021chx,Bramante:2021dyx,Zhang:2025xzi}. Rather than using large instrumented detector volumes, paleo-detectors analyze small samples of natural minerals that have resided deep underground for up to $\sim 1~\mathrm{Gyr}$, recording nuclear recoil damage tracks from rare scattering events. 

In this contribution, we summarize the results of Ref.~\cite{Theodosopoulos:2026ehn}, which explored the projected sensitivity of paleo-detectors to WIMP--nucleon interactions within the framework of a non-relativistic effective field theory (NREFT)~\cite{Fan:2010gt,Fitzpatrick:2012ix}. Sensitivities were projected for a broad range of target minerals and readout scenarios, covering the full set of NREFT operators for elastic scattering and extending to inelastic scattering with transitions to heavier WIMP states~\cite{Tucker-Smith:2001myb}. As in previous studies, the recoil track length, determined by the stopping power of nuclei in the target mineral, is used as a proxy for recoil energy.

Paleo-detectors are projected to provide sensitivity complementary to conventional direct-detection experiments and exceed existing constraints at both low and high WIMP masses. At low masses, their sensitivity benefits from low effective recoil thresholds and light target nuclei, while at high masses it is driven by the large exposures achievable over geological timescales. Relative to current limits, paleo-detectors can improve sensitivity to elastic WIMP scattering by up to several orders of magnitude at $\mathcal{O}(\mathrm{GeV})$ masses, and by up to an order of magnitude at $\mathcal{O}(\mathrm{TeV})$ masses, including inelastic scenarios with mass splittings $\lesssim 50~\mathrm{keV}$.

This paper is organized as follows. 
In Section~\ref{nreft}, we present the non-relativistic effective field theory framework for WIMP--nucleon interactions. 
Sections~\ref{DMpaleo} and~\ref{background} describe the expected nuclear recoil signatures from WIMP interactions and the relevant background sources, respectively. 
In Section~\ref{sensitivity}, we outline the statistical methodology used to derive sensitivity projections and present results for the full set of NREFT operators in comparison with current experimental constraints. 
We conclude in Section~\ref{conclusion} with a summary of our findings and a discussion of future prospects.

\section{Non-Relativistic Effective Field Theory} \label{nreft}

A complete non-relativistic effective field theory (NREFT) description of elastic WIMP--nucleon interactions was developed in Refs.~\cite{Fan:2010gt,Fitzpatrick:2012ix}. In this framework, WIMP--nucleon interactions relevant for direct-detection experiments are expressed in terms of a finite set of effective operators defined in the non-relativistic limit.
 
Imposing Galilean invariance and momentum conservation, all elastic interactions can be constructed from four Hermitian operators
\begin{equation}
     i\frac{\vec{q}}{m_N},\quad \vec{v}^{\perp} \equiv \vec{v} + \frac{\vec{q}}{2\mu},\quad \vec{S}_{\chi},\quad \vec{S}_{N},\label{basis}
\end{equation}
where $\vec{q}$ is the momentum transfer, $m_N$ the nucleon mass, $\mu_N$ the WIMP--nucleon reduced mass, $\vec{v}^{\perp}$ the component of the WIMP–nucleon relative velocity ($\vec{v}$) perpendicular to $\vec{q}$, and $\vec{S}_{\chi}$ and $\vec{S}_{N}$ the spins of the WIMP and nucleon, respectively. Truncating the operator expansion at second order in $\vec{q}$, the most general non-relativistic interaction Lagrangian can be written as
\begin{equation}
\mathcal{L}_{\rm int} = \sum_{N =n,p} \sum_i c_i^{(N)}  \mathcal{O}_i \chi^+ \chi^- N^+ N^-,
\end{equation}
where $\mathcal{O}_i$ are the linearly independent, dimensionless NREFT operators~\cite{Fitzpatrick:2012ix,Anand:2013yka}. The operators relevant for elastic scattering are listed in Table~\ref{tab:NREFT_operators}. The coefficients $c_i^{p}$ and $c_i^{n}$ denote the effective couplings to protons and neutrons, respectively. Throughout this work, we assume elastic nucleon scattering ($N^{+}\equiv N^{-}$) and perform all calculations in the isoscalar basis, $c_i^{p}=c_i^{n}$, enabling a direct comparison with limits from XENON100~\cite{XENON:2017fdd}, LUX--ZEPLIN~\cite{LZ:2023lvz}, and PandaX--II~\cite{PandaX-II:2018woa}. The operators $\mathcal{O}_{1}$ and $\mathcal{O}_{4}$ correspond to the standard spin-independent and spin-dependent interactions, respectively.
\begin{table}[t]
\centering
\small
\setlength{\tabcolsep}{6pt}   
\renewcommand{\arraystretch}{0.9} 
\captionsetup{justification=raggedright,singlelinecheck=false}
\begin{tabular}{l l}
\hline\hline
\multicolumn{2}{c}{NREFT operators relevant for elastic WIMP--nucleon scattering} \\
\hline
$\mathcal{O}_1 = 1_\chi 1_N$ 
& $\mathcal{O}_9 = i\, \vec{S}_\chi \cdot ( \vec{S}_N \times \vec{q}/m_N )$ \\

$\mathcal{O}_3 = i\, \vec{S}_N \cdot ( \vec{q}/m_N \times \vec{v}^{\perp} )$ 
& $\mathcal{O}_{10} = i\, \vec{S}_N \cdot \vec{q}/m_N$ \\

$\mathcal{O}_4 = \vec{S}_\chi \cdot \vec{S}_N$ 
& $\mathcal{O}_{11} = i\, \vec{S}_\chi \cdot \vec{q}/m_N$ \\

$\mathcal{O}_5 = i\, \vec{S}_\chi \cdot ( \vec{q}/m_N \times \vec{v}^{\perp} )$ 
& $\mathcal{O}_{12} = \vec{S}_\chi \cdot ( \vec{S}_N \times \vec{v}^{\perp} )$ \\

$\mathcal{O}_6 = ( \vec{S}_\chi \cdot \vec{q}/m_N )
                 ( \vec{S}_N \cdot \vec{q}/m_N )$
& $\mathcal{O}_{13} = i ( \vec{S}_\chi \cdot \vec{v}^{\perp} )
                       ( \vec{S}_N \cdot \vec{q}/m_N )$ \\

$\mathcal{O}_7 = \vec{S}_N \cdot \vec{v}^{\perp}$ 
& $\mathcal{O}_{14} = i ( \vec{S}_\chi \cdot \vec{q}/m_N )
                       ( \vec{S}_N \cdot \vec{v}^{\perp} )$ \\

$\mathcal{O}_8 = \vec{S}_\chi \cdot \vec{v}^{\perp}$ 
& $\mathcal{O}_{15} = - ( \vec{S}_\chi \cdot \vec{q}/m_N )
                       [ ( \vec{S}_N \times \vec{v}^{\perp} )
                       \cdot \vec{q}/m_N ]$ \\
\hline\hline
\end{tabular}
\caption{Hermitian and Galilean invariant operators defining the non-relativistic effective theory of WIMP--nucleon elastic interactions~\cite{Fan:2010gt,Fitzpatrick:2012ix}. The operators $\mathcal{O}_{1}$ and $\mathcal{O}_{4}$ correspond to canonical spin-independent (SI) and spin-dependent (SD) interactions, respectively. Operator $\mathcal{O}_{2}$ is quadratic in $\vec{v}^{\perp}$ and $\mathcal{O}_{16}$ is a linear combination of $\mathcal{O}_{12}$ and $\mathcal{O}_{15}$ and are therefore not considered here~\cite{Anand:2013yka}.}
\label{tab:NREFT_operators}
\end{table}

In addition to elastic scattering ($\chi^-\equiv\chi^+$), we also consider inelastic WIMP--nucleon scattering, in which the incoming and outgoing dark matter states differ in mass~\cite{Tucker-Smith:2001myb}. The associated mass splitting,
$\delta_m \equiv m_{\chi,\mathrm{out}} - m_{\chi,\mathrm{in}} > 0$,
introduces a kinematic threshold that modifies the scattering condition. Energy conservation implies~\cite{Barello:2014uda}
\begin{equation}
  \delta_m + \vec{v}\!\cdot \vec{q} + \frac{|\vec{q}|^2}{2\mu_N} = 0,
\end{equation}
which can be incorporated into the NREFT formalism by redefining the perpendicular velocity as
\begin{equation}
\vec v_{\text{inel}}^{\perp} \equiv \vec v + \frac{\vec q}{2\mu_N} + \frac{\delta_m}{|\vec q|^2}\vec q
= \vec v^{\perp} + \frac{\delta_m}{|\vec q|^2}\vec q ~.\label{u_inelastic}
\end{equation}
The operators describing inelastic interactions are obtained from those in Table~\ref{tab:NREFT_operators} by replacing $\vec{v}^{\perp}$ with $\vec{v}_{\mathrm{inel}}^{\,\perp}$. In this contribution, we focus on non-relativistic WIMP--nucleon elastic scattering produced via the operators in Table~\ref{tab:NREFT_operators}, with additional discussion of inelastic scattering.

\section{Dark Matter Signals in Paleo-Detectors}\label{DMpaleo}

In this section, we summarize the predictions for dark matter signals in paleo-detectors obtained in Ref.~\cite{Theodosopoulos:2026ehn}. In paleo-detectors, WIMP--nucleus scattering events produce nuclear recoils that leave damage tracks in the target mineral. Since these tracks accumulate over geological timescales and their lengths are determined by the stopping power of recoiling nuclei, the track length provides a direct proxy for the nuclear recoil energy.
 
For a WIMP of mass $m_{\chi}$ scattering elastically off a target nucleus $(Z,A)$ of mass $m_T$, the differential nuclear-recoil rate in the NREFT framework~\cite{Fitzpatrick:2012ix,Anand:2013yka} reads
\begin{equation}
    \left(\frac{dR}{dE_R}\right)_{(Z,A)} = N_T \frac{\rho_\chi m_T}{32 \pi m_\chi^3 m_N^2} \left\langle \frac{1}{v} \sum_{i j} \sum_{N,N'=p,n} c_i^{(N)} c_j^{(N')} F_{ij,\;(Z,A)}^{(N,N')}\!\left(v^2, q^2\right) \right\rangle~,\label{recoilrate}
\end{equation}
where $R$ is the number of recoiling nuclei per unit exposure\footnote{In this contribution, we assume a one-to-one correspondence between the observed number of tracks and the number of recoiling nuclei; under this assumption, the rate $R$ in Eq.~\eqref{recoilrate} in consistent with the interpretation as the number of tracks per unit exposure.}, $E_{R}$ is the nuclear recoil energy, $N_T$ is the number of target nuclei per detector mass, $m_{N}$ is the nucleon mass, $\rho_{\chi}$ is the local DM density, $v$ is the WIMP speed following the Standard Halo Model velocity distribution in the laboratory frame, $F_{ij,\;(Z,A)}^{(N, N')}$ are the form factors defined in \cite{Fitzpatrick:2012ix,Anand:2013yka}, and $\langle...\rangle$ indicates average over the halo velocity distribution. For the Standard Halo Model, we adopt the conventions of Ref.~\cite{Baxter:2021pqo}. The recoil rates were computed using the open-source codes WimPyDD \cite{Jeong:2021bpl} and dmscatter \cite{Gorton:2022eed} were utilized, whose results have been verified by comparison to those of the Mathematica package DMFormFactor \cite{Anand:2013yka}.  

The damage-track length $x_T$ produced by a recoiling nucleus ($Z,A$) with recoil energy $E_R$ is obtained from
\begin{equation}
x_T(E_R) = \int_0^{E_R} dE \, \left| \frac{dE}{dx} \right|_{(Z,A)}^{-1},
\label{length}
\end{equation}
where $(dE/dx)_{(Z,A)}$ is the stopping power, evaluated using \texttt{SRIM}~\cite{ZIEGLER20101818}. Deviations between the idealized range and the observed track length are expected to be small~\cite{Drukier:2018pdy}. Tracks produced by light ions with $Z\le2$ are excluded, as they do not form stable or detectable damage tracks~\cite{Drukier:2018pdy}.

For a mineral composed of multiple nuclear species, the total differential track-production rate (per unit exposure, and per unit track length) is
\begin{equation}
    \frac{dR}{dx_T}=\sum_{(Z,A)}\xi_{(Z,A)}\left(\frac{dR}{dE_{R}}\right)_{(Z,A)}\left(\frac{dE_R}{dx_T}\right)_{(Z,A)}~,\label{dRdx}
\end{equation}
where $\xi_{(Z,A)}$ is the mass fraction of each nuclear species, and $(dR/dE_R)_{(Z,A)}$ denotes the recoil energy spectrum for a given nuclear species, as defined in Eq.~\eqref{recoilrate}. 
The factor $(dE_R/dx_T)_{(Z,A)}$ accounts for the conversion from recoil energy to track length and is evaluated using stopping power calculations from \texttt{SRIM}~\cite{ZIEGLER20101818}. Plots of the differential event rate for all NREFT operators and for various target minerals can be found in Ref.~\cite{Theodosopoulos:2026ehn}. 

Below, we present predictions for event rates in several target materials, whose chemical compositions and ${}^{238}$U concentrations are summarized in Table~\ref{tab:U238_concentration}. 
In particular, we focus on four minerals commonly found in nature: two ultrabasic rocks (UBRs), olivine and muscovite, the latter of which contains hydrogen, and two marine evaporites (MEs), halite and gypsum, with gypsum also containing hydrogen. 
These mineral classes are chosen because they typically exhibit lower ${}^{238}$U concentrations than many other geological materials. 
Both the uranium content and the presence of hydrogen in the target material play an important role in determining the expected background rates in paleo-detector experiments, as discussed in Section~\ref{background}. 
\begin{table}[ht]
\captionsetup{justification=raggedright,singlelinecheck=false}
\centering
\begin{tabular}{l l c}
\hline
\textbf{Mineral} & \textbf{Composition} & \textbf{Fiducial ${}^{238}$U Concentration [per Weight, g/g]} \\
\hline
Gypsum        & Ca(SO$_4$)$\cdot$2(H$_2$O)     & $10^{-11}$ \\
Halite        & NaCl                          & $10^{-11}$ \\
Olivine       & Mg$_{1.6}$Fe$^{2+}_{0.4}$(SiO$_4$) & $10^{-10}$ \\
Muscovite    & KAl$_2$(AlSi$_3$O$_{10}$)(F,OH)$_{2}$   & $10^{-10}$ \\
\hline
\end{tabular}
\caption{Minerals considered in this work, together with their chemical compositions and the fiducial ${}^{238}$U concentrations $(C^{238})$ assumed for radiopure samples.}
\label{tab:U238_concentration}
\end{table}

Nuclear recoil tracks recorded in paleo-detector samples can be read out using a variety of experimental techniques~\cite{Drukier:2018pdy}.
We consider two benchmark readout scenarios: (i) the high-resolution (HR) scenario assumes a sample mass of $M=10~\mathrm{mg}$ and a spatial resolution of $\sigma_x=1~\mathrm{nm}$, which may be achievable with helium ion beam microscopy in combination with pulsed-laser and fast-ion-beam ablation techniques~\cite{HILL201265,VANGASTEL20122104,Joens2013,ECHLIN20151,PFEIFENBERGER2017109,10.1116/1.5047806}, and is particularly well suited for searches for low-mass WIMPs with $m_\chi \lesssim 10~\mathrm{GeV}/c^{2}$~\cite{Baum:2021jak}. (ii) the high-exposure (HE) scenario assumes $M=100~\mathrm{g}$ and $\sigma_x=15~\mathrm{nm}$, potentially realizable using small-angle X-ray scattering tomography at synchrotron facilities~\cite{RODRIGUEZ2014150,Schaff2015,Holler2014}, and is better suited for probing heavier WIMPs with $m_\chi \gtrsim 10~\mathrm{GeV}/c^{2}$~\cite{Baum:2021jak}.

The main result presented in this section is the predicted binned track--length spectrum, defined as the number of tracks arising from WIMP--nucleus interactions binned by track length, which is the primary observable in a paleo-detector.  
For a mineral sample of mass $M$ and age $t_{\mathrm{age}}$, the expected number of tracks in the $k$-th bin, corresponding to track lengths $x_T \in [x_{T,k}^{\min}, x_{T,k}^{\max}]$, reads
\begin{equation}
    \mathcal{N}_k = M \times t_{\mathrm{age}} \int dx_T' \, W\!\left(x_T'; x_{T,k}^{\min}, x_{T,k}^{\max}\right) \frac{dR}{dx_T}(x_T')~.\label{bin}
\end{equation}
The finite readout resolution is modeled by assuming that the probability of reconstructing a track length $x_T$ from a true length $x_T'$ follows a Gaussian distribution with variance $\sigma_x^2$, corresponding to the square of the spatial resolution. 
This leads to the window function 
\begin{equation}
    W\!\left(x_T'; x_{T,k}^{\min}, x_{T,k}^{\max}\right) = \frac{1}{2} \left[ \operatorname{erf} \left( \frac{x_{T}' - x_{T,k}^{\min}}{\sqrt{2}\,\sigma_{x_T}} \right) - \operatorname{erf} \left( \frac{x_{T}' - x_{T,k}^{\max}}{\sqrt{2}\,\sigma_{x_T}} \right) \right]~.
\end{equation}
To avoid artificially enhancing sensitivity, we exclude all tracks with true lengths $x_T' < \sigma_x/2$. 
All numerical results use 100 logarithmic bins between $\sigma_x/2$ and $1000~\mathrm{nm}$. 

\begin{figure}
    \captionsetup{justification=raggedright,singlelinecheck=false}
    \centering
    \includegraphics[width=0.48\textwidth]{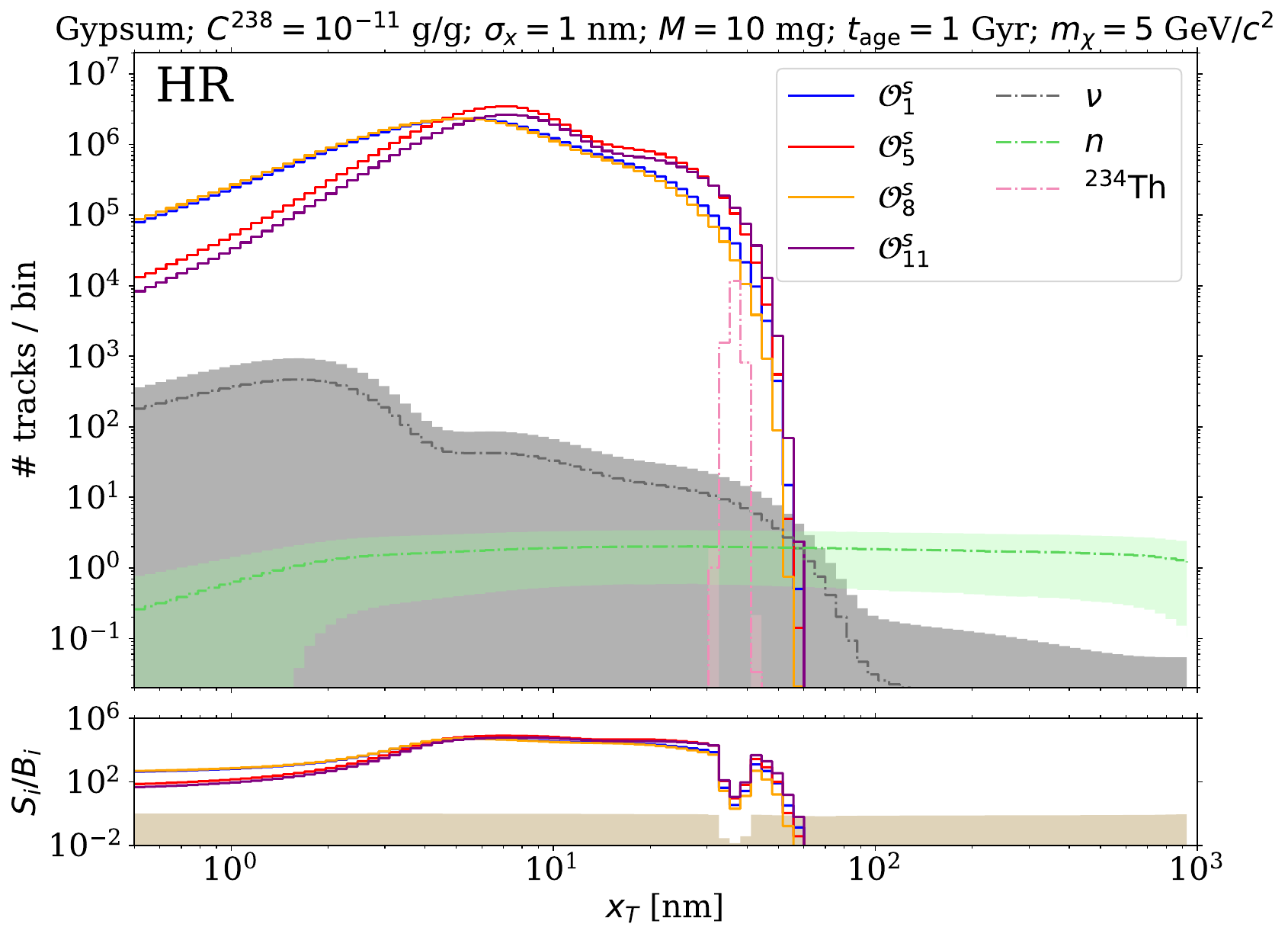}
    \includegraphics[width=0.49\textwidth]{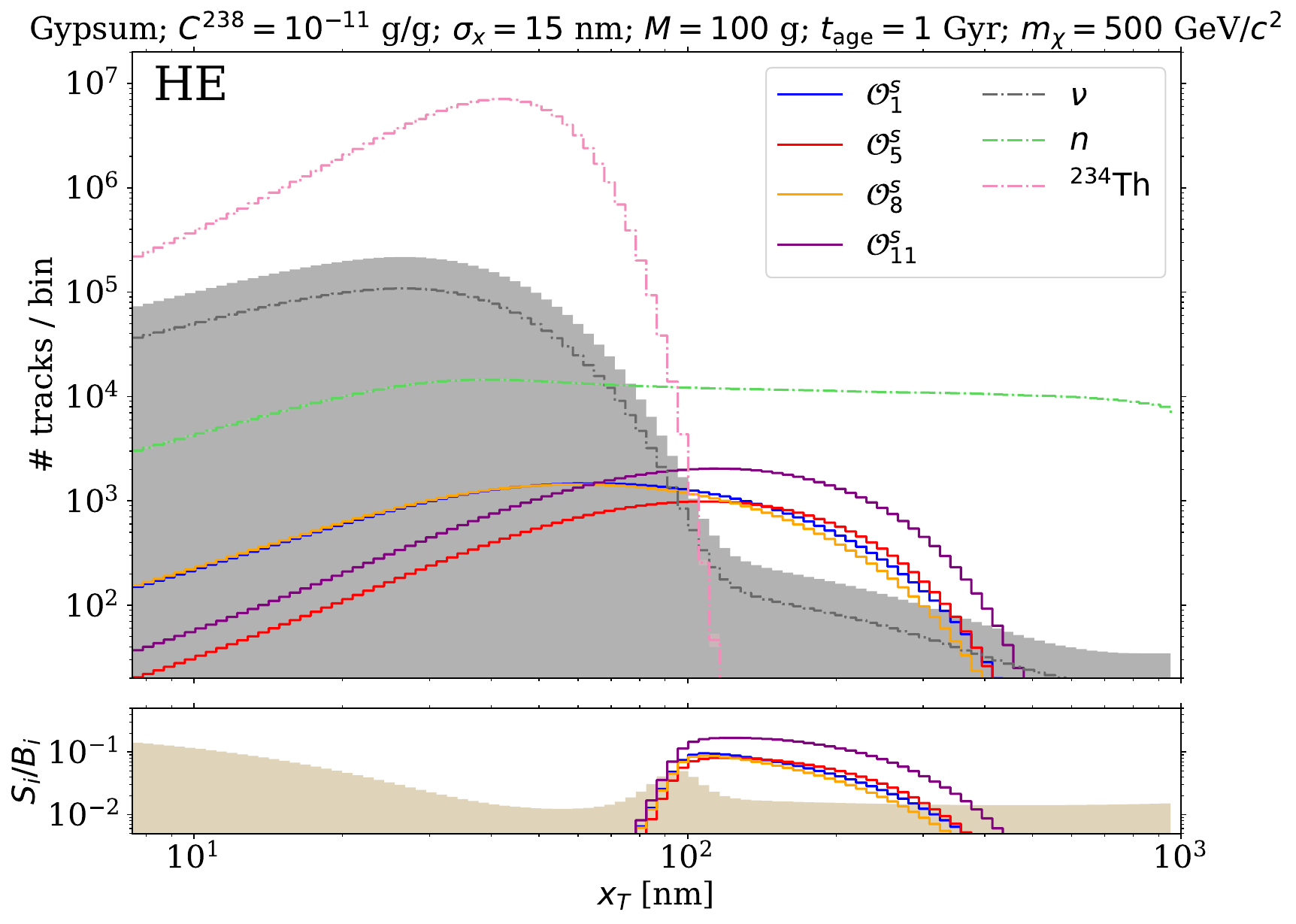}
    \caption{Upper panels: Track--length spectra (number of tracks binned by track length)
    for the spin–independent NREFT operators $\mathcal{O}^{s}_{1}$, $\mathcal{O}^{s}_{5}$, $\mathcal{O}^{s}_{8}$, and $\mathcal{O}^{s}_{11}$ in gypsum, assuming  the case of elastic, isoscalar interactions ($c^{p}=c^{n}$). Left panel: HR scenario, with a DM mass of $m_{\chi}=5\ \mathrm{GeV}/c^{2}$. For the normalization of the DM track production spectra, we set the NREFT coupling constants to be equal to the upper limits from the SuperCDMS experiment \cite{SuperCDMS:2022crd}. Right panel: HE scenario, with $m_{\chi}=500\ \mathrm{GeV}/c^{2}$, and coupling constants compatible with the upper limits from the LUX--ZEPLIN experiment \cite{LZ:2023lvz}. For comparison, background spectra induced by neutrinos ($\nu$), radiogenic neutrons ($n$), and ${}^{238}\text{U}\to{}^{234}\text{Th}+\alpha$ recoils (${}^{234}\text{Th}$) are also shown; see Sec.~\ref{background}. Here, we include shaded bands around background components, representing the combined statistical (Poisson) and systematic uncertainties in the background predictions. Bottom panels: Colored lines show the ratio of signal events ($S_{i}$) to background events ($B_{i}$) in each bin. The upper edge of the sand–colored band indicates the relative uncertainty ($\delta B_i/B_i$) for the number of background events per bin. Figures adapted from Ref.~\cite{Theodosopoulos:2026ehn}.}
    \label{fig:Spectrum_binned_SI}
\end{figure}

\begin{figure}
    \captionsetup{justification=raggedright,singlelinecheck=false}
    \centering
    \includegraphics[width=0.48\textwidth]{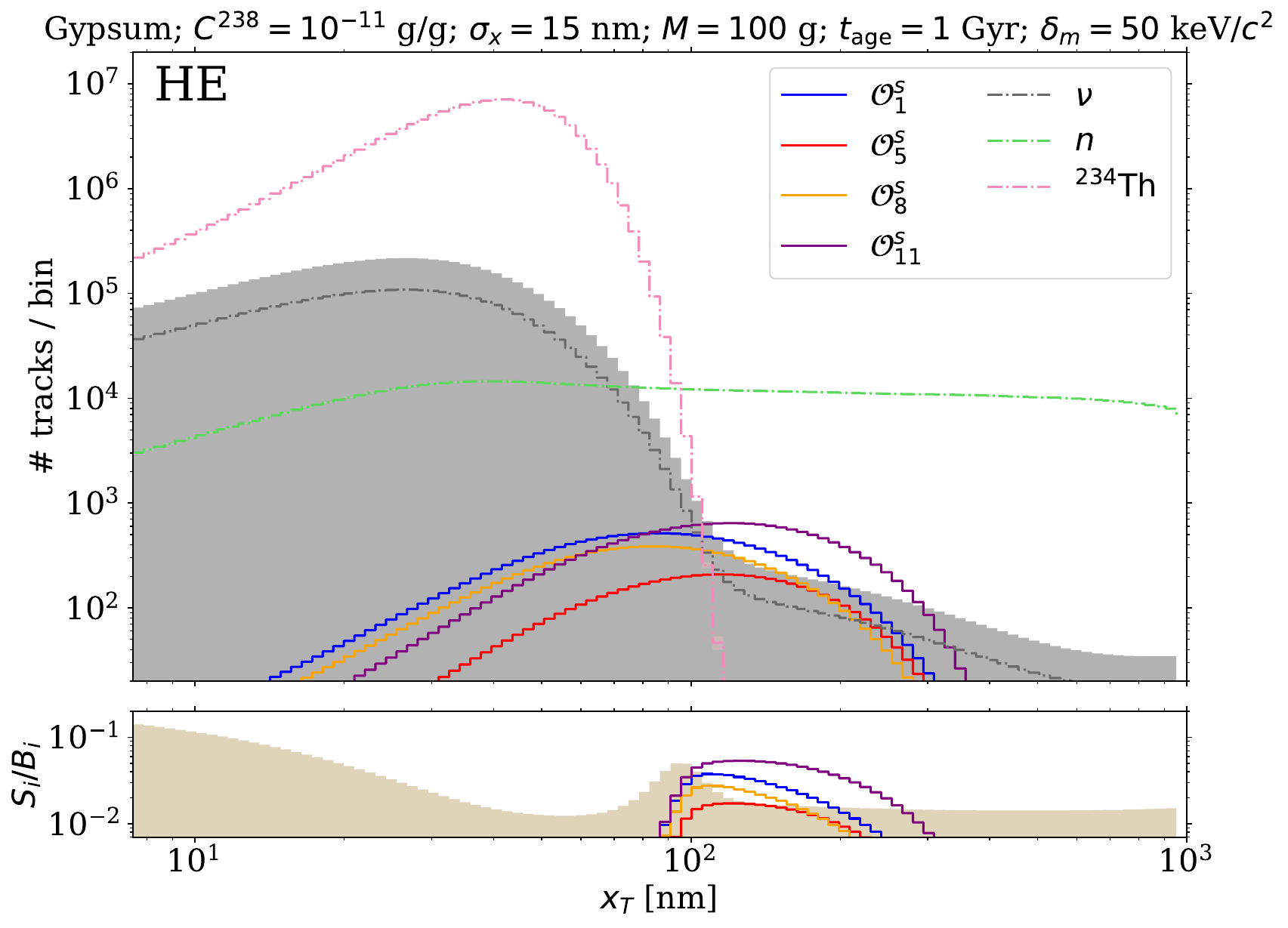}
    \includegraphics[width=0.485\textwidth]{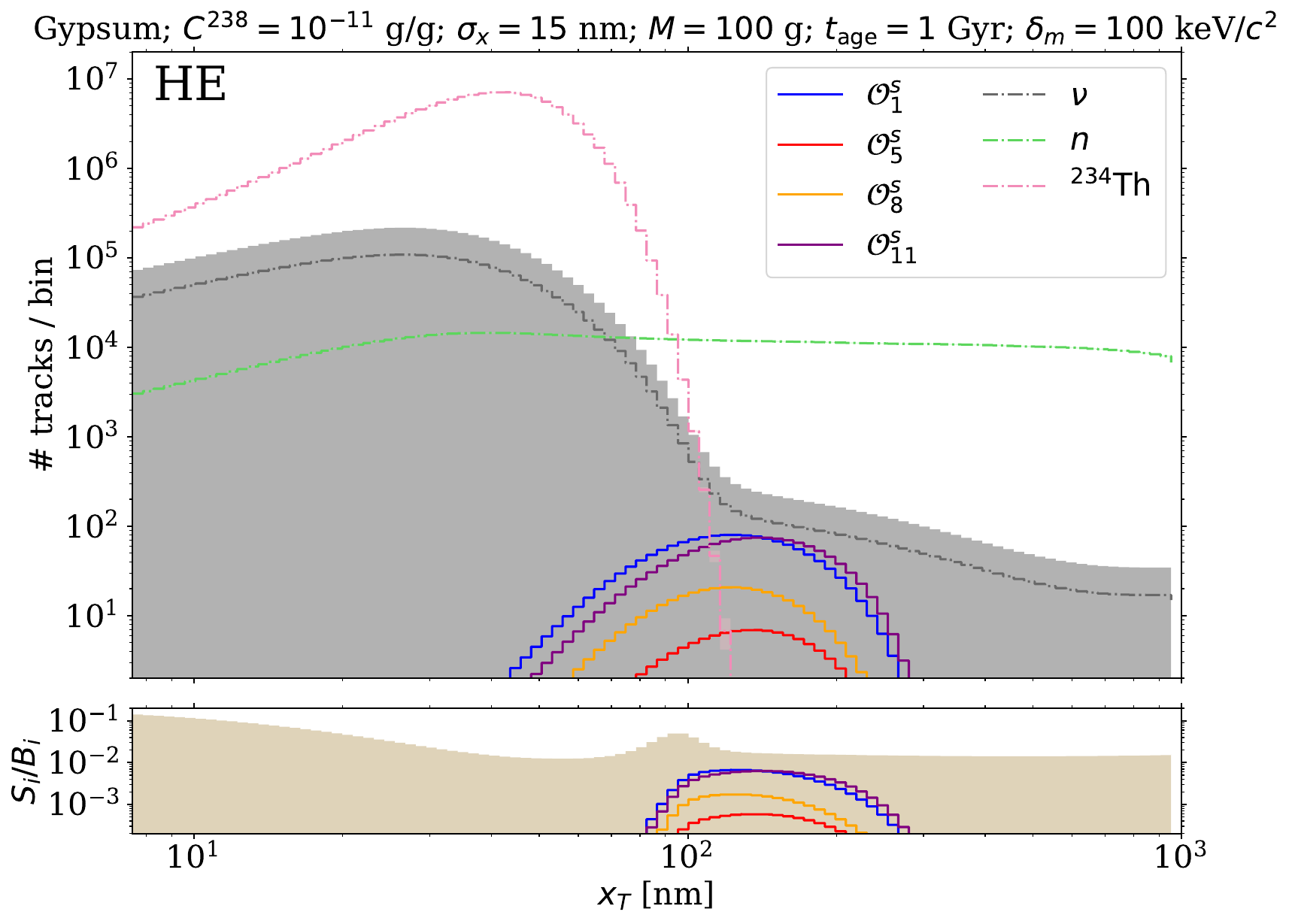}
    \caption{Upper panels: Track--length spectra (number of tracks binned by track length) for the spin–independent NREFT operators $\mathcal{O}^{s}_{1}$, $\mathcal{O}^{s}_{5}$, $\mathcal{O}^{s}_{8}$, and $\mathcal{O}^{s}_{11}$
    in gypsum for the case of inelastic, isoscalar interactions ($c^{p}=c^{n}$), and a DM mass of $m_{\chi}=500\ \mathrm{GeV}/c^{2}$, in the HE scenario. In the left panel, the mass splitting is $\delta_{m}=50\ \mathrm{keV}/c^{2}$. In the right panel, the mass splitting is $\delta_{m}=100\ \mathrm{keV}/c^{2}$. In both panels, the values for the coupling constants are chosen to be compatible with the upper limits for inelastic WIMP--nucleon scattering from the LUX--ZEPLIN experiment \cite{LZ:2023lvz}. For comparison, background spectra induced by neutrinos ($\nu$), radiogenic neutrons ($n$), and ${}^{238}\text{U}\to{}^{234}\text{Th}+\alpha$ recoils (${}^{234}\text{Th}$) are also shown; see Sec.~\ref{background}. Here, we include shaded bands around background components, representing the combined statistical (Poisson) and systematic uncertainties in the background predictions. Bottom panels: Ratio of signal events ($S_{i}$) to background events ($B_{i}$) in each bin (colored lines). The sand–colored band indicates the relative uncertainty ($\delta B_i/B_i$) for the number of background events per bin. Figures adapted from Ref.~\cite{Theodosopoulos:2026ehn}. }
    \label{fig:Spectrum_binned_SI_inelastic}
\end{figure}

Figs.~\ref{fig:Spectrum_binned_SI} and~\ref{fig:Spectrum_binned_SI_inelastic} show the resulting binned track-length spectra in gypsum for the spin-independent NREFT operators $\mathcal{O}_1^s$, $\mathcal{O}_5^s$, $\mathcal{O}_8^s$, and $\mathcal{O}_{11}^s$, for elastic and inelastic scattering, respectively. In Fig.~\ref{fig:Spectrum_binned_SI}, the HR (HE) scenario is shown for representative low-mass (high-mass) WIMP benchmarks. In the HR and HE scenarios, the NREFT coupling constants are chosen to be consistent with the 95\% Bayesian credible region of the two-dimensional marginalized posterior distribution reported by SuperCDMS~\cite{SuperCDMS:2022crd} and with the 90\% confidence level exclusion limits on the isoscalar WIMP--nucleon NREFT coupling constants ($c^{p}=c^{n}$) for elastic scattering reported by the LUX--ZEPLIN experiment~\cite{LZ:2023lvz}, respectively. Background contributions (see Section~\ref{background}) from neutrinos, radiogenic neutrons, and ${}^{234}$Th recoils are included, together with their statistical and systematic uncertainties (shaded bands).

The bottom panels of Fig.~\ref{fig:Spectrum_binned_SI} display, as colored curves, the ratio of signal events $S_i$ to background events $B_i$ in each track--length bin.
The sand-colored region represents the relative uncertainty in the background events per bin. 
Specifically, the upper edge of the sand-colored region corresponds to the ratio of the total (statistical $+$ systematic) uncertainty in the background events, $\delta B_i$, to the expected number of background events, $B_i$. 
Consequently, the ratio of the colored curves ($S_i/B_i$) to the upper edge of the sand-colored region ($\delta B_i/B_i$) provides an estimate of the signal-to-noise ratio ($S_i/\delta B_i$) in each bin. This signal-to-noise ratio offers a qualitative measure of the sensitivity of paleo-detectors: when it exceeds unity, paleo-detectors are expected to outperform the corresponding direct-detection experiment whose limits are used to normalize the track--length spectra shown in the plots.
For low-mass WIMPs, nuclear recoils are typically softer, leading to shorter damage tracks, as illustrated in the left panel. 
In this regime, achieving high track--length resolution is essential for distinguishing signal and background features in the region where the signal-to-background ratio is largest. 
By contrast, heavier WIMPs produce longer tracks on average, as shown in the right panel. 
Since existing direct-detection experiments already constrain much larger NREFT coupling constants for $m_\chi \gtrsim 10~\mathrm{GeV}/c^{2}$, probing this mass range with couplings below current limits requires the large effective exposures provided by the HE scenario.

In the inelastic case, sensitivity is limited by the relatively light target nuclei available in natural minerals. As discussed in Ref.~\cite{Theodosopoulos:2026ehn}, paleo-detectors can probe mass splittings up to $\delta_m \sim 100~\mathrm{keV}/c^2$, with sensitivity maximized for heavy WIMPs. Accordingly, Fig.~\ref{fig:Spectrum_binned_SI_inelastic} shows results for $m_\chi=500~\mathrm{GeV}/c^2$, representative mass splittings $\delta_m=50$ and $100~\mathrm{keV}/c^2$, and coupling constants consistent with the upper limits for inelastic scattering reported by the LUX--ZEPLIN experiment \cite{LZ:2023lvz}.
As shown in the top panels of Fig.~\ref{fig:Spectrum_binned_SI_inelastic}, inelastic scattering leads to suppressed signal spectra in the HE scenario. As illustrated in the bottom panels of Fig.~\ref{fig:Spectrum_binned_SI_inelastic}, for the $\mathcal{O}_{1}^{s}$ and $\mathcal{O}_{11}^{s}$ operators, which are independent of $\vec{v}_{\mathrm{inel}}^{\,\perp}$, the signal-to-background ratio remains clearly distinguishable from the background uncertainty. In contrast, for the $\mathcal{O}_{5}^{s}$ and $\mathcal{O}_{8}^{s}$ operators, which depend on $\vec{v}_{\mathrm{inel}}^{\,\perp}$, the signal-to-background ratio is comparable to the background uncertainty.

In this section, we have presented predictions for the binned track length spectrum, which constitutes the experimentally observable quantity (see Figs.~\ref{fig:Spectrum_binned_SI} and~\ref{fig:Spectrum_binned_SI_inelastic}). Results are shown for the spin-independent NREFT operators $\mathcal{O}_{1}^{s}$, $\mathcal{O}_{5}^{s}$, $\mathcal{O}_{8}^{s}$, and $\mathcal{O}_{11}^{s}$, considering both elastic and inelastic scattering.

\section{Backgrounds in paleo-detector dark matter searches}\label{background}

The backgrounds relevant for DM searches with paleo-detectors are similar to those encountered in conventional direct-detection experiments, but their relative importance differs due to the extremely large effective exposures achievable over geological timescales. The dominant background sources can be grouped into three categories: cosmogenic backgrounds, astrophysical neutrinos, and radiogenic processes. Detailed discussions can be found in Ref.~\cite{Drukier:2018pdy}.

Cosmogenic backgrounds are strongly suppressed by selecting minerals that were shielded by a substantial overburden during the period in which damage tracks were accumulated. Since only gram-scale samples are required for paleo-detector analyses, materials can be obtained from boreholes at depths exceeding those of conventional underground laboratories.
At depths of approximately $5~\mathrm{km}$, the cosmogenic muon--induced neutron flux is of order
$\mathcal{O}(10^{2})~\mathrm{cm}^{-2}\,\mathrm{Gyr}^{-1}$~\cite{Mei:2005gm}, making this background negligible for DM searches. 

Astrophysical neutrinos produce nuclear recoils in paleo-detector samples and constitute an irreducible background. We include neutrinos from the Sun, from core-collapse supernovae, and from cosmic-ray interactions in the Earth’s atmosphere~\cite{OHare:2020lva}. Due to their long integration times, paleo-detectors are sensitive both to the diffuse supernova neutrino background (DSNB) and to neutrinos from supernovae within the Milky Way, whose rate is estimated to be $2$--$3$ per century~\cite{Cappellaro:2003eg}. The DSNB and Galactic supernova contributions are computed following Ref.~\cite{Baum:2019fqm}. The relative importance of neutrino backgrounds depends on the track length $x_T$: solar neutrinos dominate at short track lengths ($x_T\lesssim100~\mathrm{nm}$), supernova neutrinos dominate at intermediate lengths ($10~\mathrm{nm}\lesssim x_T\lesssim500~\mathrm{nm}$), and atmospheric neutrinos dominate at larger track lengths.

Radiogenic backgrounds arise from trace amounts of radioactive isotopes present in natural minerals. The dominant contribution comes from $^{238}\mathrm{U}$. Typical crustal rocks contain uranium concentrations of order
$C^{238}\sim10^{-6}\,\mathrm{g/g}$, which would generate prohibitively large backgrounds for DM searches. To suppress these backgrounds, paleo-detector studies focus on materials with exceptionally low uranium content, such as ultra-basic rocks (UBRs), originating from the Earth’s mantle, and marine evaporites (MEs), formed from evaporated seawater (see Ref.~\cite{Drukier:2018pdy} and the appendix of Ref.~\cite{Baum:2019fqm}). Following previous work~\cite{Baum:2018tfw,Drukier:2018pdy,Edwards:2018hcf,Baum:2019fqm,Theodosopoulos:2026ehn,Baum:2021jak}, we adopt benchmark uranium concentrations of
$C^{238}=10^{-10}\,\mathrm{g/g}$ for UBRs and $C^{238}=10^{-11}\,\mathrm{g/g}$ for MEs. 
The dominant radiogenic background processes are $\alpha$ decays and neutron production via spontaneous fission and $(\alpha,n)$ reactions~\cite{Drukier:2018pdy,Baum:2021jak}. The dominant source of isolated nuclear recoil backgrounds arises from the initial decay
$^{238}\mathrm{U}\to{}^{234}\mathrm{Th}+\alpha$, which yields a $^{234}\mathrm{Th}$ recoil with a fixed energy of $72~\mathrm{keV}$. These events form a monochromatic population in track--length space with number density
$n(^{234}\mathrm{Th})\simeq10^{6}~\mathrm{g}^{-1}\,(C^{238}/10^{-11}~\mathrm{g/g})$~\cite{Drukier:2018pdy} and can therefore be readily identified. 
Additional radiogenic backgrounds arise from neutrons produced in spontaneous fission and $(\alpha,n)$ reactions. These neutrons generate a broad spectrum of nuclear recoil tracks. We compute neutron spectra using SOURCES-4A~\cite{sources4a1999} and recoil spectra using neutron--nucleus cross sections from the TENDL-2017 library~\cite{KONING20122841}, accessed via JANIS4.0~\cite{SOPPERA2014294}. Neutron-induced backgrounds are strongly suppressed in hydrogen-bearing minerals, as neutrons lose a large fraction of their energy in single collisions with hydrogen, making such materials particularly favorable for paleo-detector searches~\cite{Baum:2018tfw,Drukier:2018pdy}.

\section{Projected Sensitivity of Paleo-Detectors}\label{sensitivity}

In this section, we summarize the projected sensitivity of paleo-detectors to WIMP--nucleon interactions, following the results originally obtained in Ref.~\cite{Theodosopoulos:2026ehn}. Because of the extremely large effective exposures achievable with paleo-detectors, any realistic mineral sample is expected to contain a very large number of background events. As a consequence, DM searches with paleo-detectors cannot rely on background-free signal regions, but instead require a spectral analysis of the recoil track--length distribution. 
The search strategy consists of testing whether the observed track--length spectrum is better described by a background-only hypothesis or by a combined background-plus-signal model. To this end, we employ a profile-likelihood--ratio approach, which allows us to project the sensitivity of paleo-detectors to a DM signal using the full spectral information.

The expected binned track--length spectrum, i.e.~the number of tracks in the $k^{\rm th}$ bin, including both background and DM signal contributions, is given by
\begin{equation}
\begin{aligned}
    \mathcal{N}_k(\vec{\theta}; c_j, m_{\chi}) &=
    \mathcal{N}_{\nu,k}^{\mathrm{sol}}\!\left(\Phi_{\nu}^{\mathrm{sol}}\right)
    + \mathcal{N}_{\nu,k}^{\mathrm{DSNB}}\!\left(\Phi_{\nu}^{\mathrm{DSNB}}\right)
    +\mathcal{N}_{\nu,k}^{\mathrm{GSNB}}\!\left(\Phi_{\nu}^{\mathrm{GSNB}}\right)
    + \mathcal{N}_{\nu,k}^{\mathrm{atm}}\!\left(\Phi_{\nu}^{\mathrm{atm}}\right) \\
    &\quad + \mathcal{N}_{\mathrm{rad},k}^{^{234}\mathrm{Th}}\!\left(C^{238}\right)
    + \mathcal{N}_{\mathrm{rad},k}^{n}\!\left(C^{238}\right)
    + \mathcal{N}_{\mathrm{DM},k}\!\left(c_j, m_{\chi}\right) ~,
\end{aligned}
\end{equation}
where $\mathcal{N}_{s,k}$ denotes the binned track--length spectra for the various signal and background components, computed according to Eq.~(\ref{bin}). The quantities $\Phi_{\nu}$ represent the neutrino fluxes, $C^{238}$ is the $^{238}\mathrm{U}$ concentration in the target mineral, $c_j$ is the coupling constant of a given NREFT interaction, and $m_{\chi}$ is the DM mass.
The set of nuisance parameters entering the likelihood is $\vec{\theta}=\left\{M, t_{\text{age}},\Phi_{\nu}^{\mathrm{sol}},\Phi_{\nu}^{\mathrm{DSNB}},\Phi_{\nu}^{\mathrm{GSNB}},\Phi_{\nu}^{\mathrm{atm}},C^{238}\right\}$,
where $M$ and $t_{\mathrm{age}}$ denote the mass and age of the mineral sample, respectively. The neutrino-induced backgrounds arise from solar neutrinos ($\mathcal{N}_{\nu}^{\mathrm{sol}}$), the diffuse supernova neutrino background ($\mathcal{N}_{\nu}^{\mathrm{DSNB}}$), neutrinos from Galactic supernovae ($\mathcal{N}_{\nu}^{\mathrm{GSNB}}$), and atmospheric neutrinos ($\mathcal{N}_{\nu}^{\mathrm{atm}}$). Radiogenic backgrounds consist of two contributions: isolated recoils of $^{234}\mathrm{Th}$ nuclei originating from the initial $\alpha$ decay of $^{238}\mathrm{U}$ during the recording period, and nuclear recoils induced by radiogenic neutrons.

The DM signal contribution $\mathcal{N}_{\mathrm{DM},k}(c_j, m_{\chi})$ depends on the NREFT coupling constant $c_j$ and the DM mass $m_{\chi}$. For NREFT operators that depend on the DM spin $\vec{S}_{\chi}$, the signal also depends on the spin of the DM particle, which we take to be $1/2$ throughout this work. All contributions scale linearly with the sample mass $M$, while all components except the $^{234}\mathrm{Th}$ contribution are also proportional to the sample age $t_{\mathrm{age}}$.

To facilitate a direct comparison with existing limits and projected sensitivities from conventional direct-detection experiments, we compute projected $90\%$ confidence level upper limits on the NREFT coupling constants as a function of the DM mass $m_{\chi}$. Since paleo-detectors measure the number of nuclear recoil tracks accumulated over geological timescales, they are effectively counting experiments, and the number of events in each track--length bin is assumed to follow Poisson statistics. Given a data set $\mathbf{D}$ and a parameter set $\{\vec{\theta}, c_j, m_{\chi}\}$, the Poisson log-likelihood is
\begin{equation}
    \ln \mathcal{L}_{\text{Poisson}}\bigl(\mathbf{D}\,|\,\vec{\theta}; c_j, m_{\chi}\bigr)
    = \sum_{i} \left[ D_{i} \ln \mathcal{N}_{i}\bigl(\vec{\theta}; c_j, m_{\chi}\bigr)
    - \mathcal{N}_{i}\bigl(\vec{\theta}; c_j, m_{\chi}\bigr) \right]~,
\end{equation}
where $\mathcal{N}_{i}$ denotes the expected number of events in the $i^{\rm th}$ bin. Constant terms independent of the parameters have been omitted, as they cancel in likelihood ratios.

The nuisance parameters $\vec{\theta}$ are constrained by independent measurements. Neutrino fluxes are informed by existing neutrino experiments and theoretical calculations, while the mass, age, and uranium content of the mineral sample can be determined experimentally. In a frequentist framework, these external constraints are incorporated by including Gaussian constraints in the likelihood,
\begin{equation}
\ln \mathcal{L}_{\text{ext. const.}}\left(\vec{\theta}\right)
= \sum_{j} \left[ -\frac{1}{2}
\left( \frac{\vec{\theta}_{j} - \bar{\vec{\theta}}_{j}}{\ell_{j}\,\bar{\vec{\theta}}_{j}} \right)^{2} \right]~,
\end{equation}
where $\bar{\theta}_{j}$ denotes the central value inferred from ancillary data and $\ell_{j}$ is the corresponding relative uncertainty. The values of $\bar{\theta}_{j}$ and $\ell_{j}$ used for this analysis can be found in Ref.~\cite{Theodosopoulos:2026ehn}.
The full likelihood used in the analysis is then given by
\begin{equation}
\ln \mathcal{L}\bigl(\mathbf{D}\,|\,\vec{\theta}; c_j, m_{\chi}\bigr)
= \ln \mathcal{L}_{\text{Poisson}}\bigl(\mathbf{D}\,|\,\vec{\theta}; c_j, m_{\chi}\bigr)
+ \ln \mathcal{L}_{\text{ext. const.}}(\vec{\theta}).
\end{equation}
To estimate the projected sensitivity of paleo-detectors, we employ the profile-likelihood--ratio test statistic~\cite{Cowan:2010js},
\begin{equation}
q(c; m_{\chi})
= -2 \ln \left[
\mathcal{L}( \mathbf{D} \big| \hat{\hat{\vec{\theta}}}; c_j, m_{\chi} )\!/
 \mathcal{L}( \mathbf{D} \big| \hat{\vec{\theta}}; \hat{c}_j, m_{\chi})
\right]~,
\end{equation}
where $\hat{\hat{\vec{\theta}}}$ denotes the values of the nuisance parameters that maximize the likelihood for fixed $c_j$ and $m_{\chi}$ in the numerator. In the denominator, the likelihood is maximized with respect to both $\vec{\theta}$ and $c_j$, yielding the best-fit values $\hat{\vec{\theta}}$ and $\hat{c}_j$ for a fixed DM mass. We can obtain projected exclusion limits using the \emph{Asimov} data set~\cite{Cowan:2010js}, corresponding to the background-only hypothesis, 
\begin{equation}
D_i = \mathcal{N}_{i}\left(\bar{\vec{\theta}}; c_j = 0 \right),
\end{equation}
where $\bar{\vec{\theta}}$ denotes a fiducial choice of nuisance parameters. For a given $m_{\chi}$, the $90\%$ confidence level exclusion limit is defined as the smallest value of $c_j$ satisfying
$q(c_j; m_{\chi}) \ge q_{\mathrm{crit}} = 2.71$.
According to Wilks’ theorem~\cite{Wilks:1938dza}, the test statistic $q$ follows a $\chi^{2}$ distribution in the asymptotic limit; for one degree of freedom, $q=2.71$ corresponds to a $p$-value of $0.1$.

\begin{figure}
    \captionsetup{justification=raggedright,singlelinecheck=false}
    \centering
    \begin{subfigure}{0.32\textwidth}
    \includegraphics[width=\linewidth]{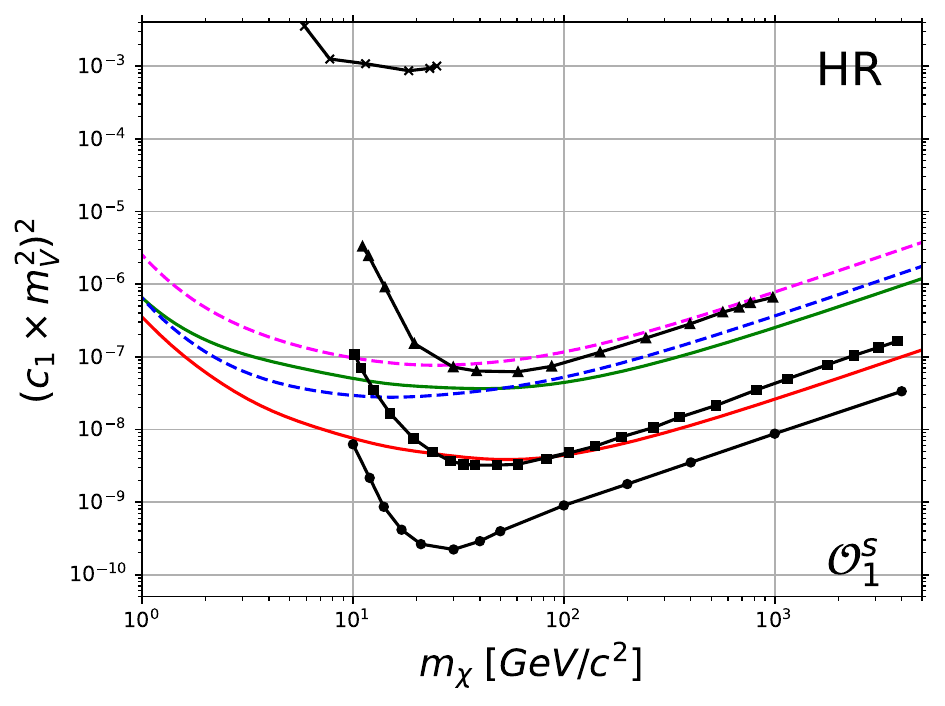}
    \end{subfigure}
    \begin{subfigure}{0.32\textwidth}
    \includegraphics[width=\linewidth]{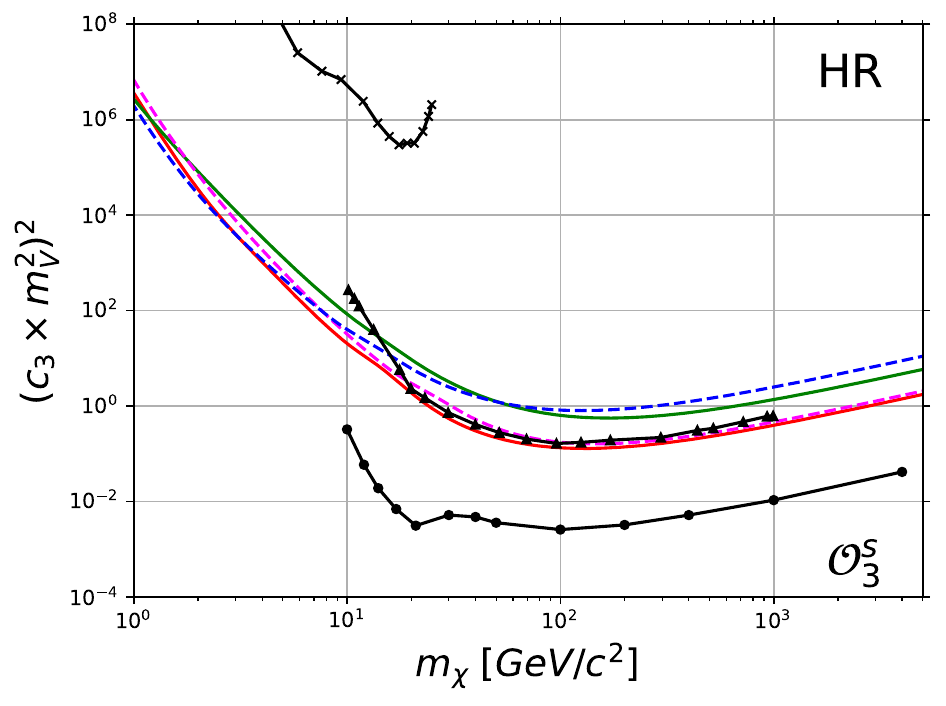}
    \end{subfigure}
    \begin{subfigure}{0.32\textwidth}
    \includegraphics[width=\linewidth]{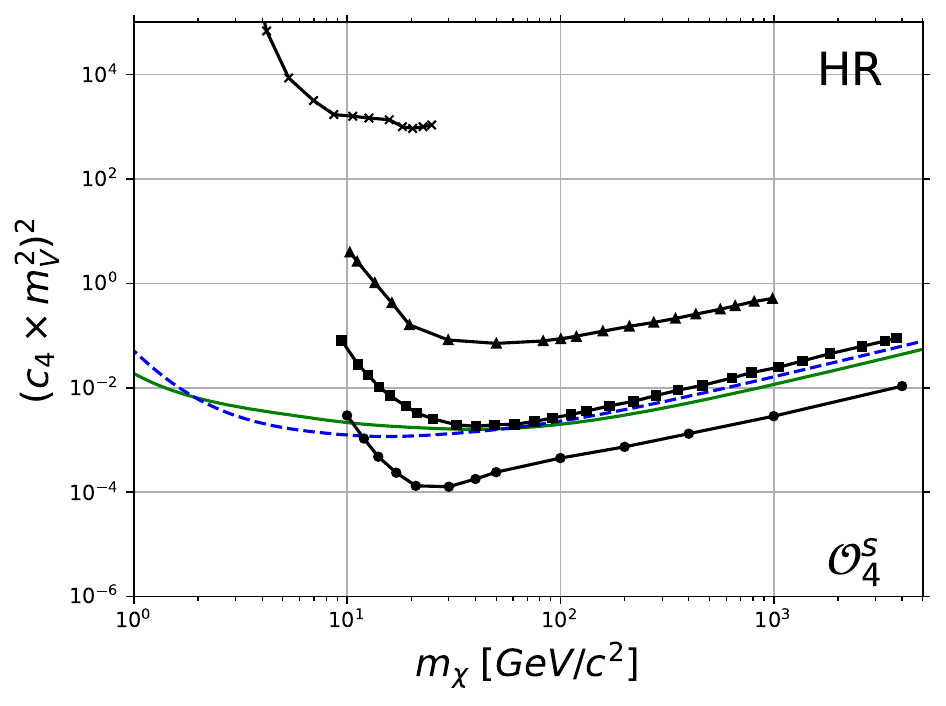}
    \end{subfigure}

    \vspace{0.1ex}

    \begin{subfigure}{0.32\textwidth}
    \includegraphics[width=\linewidth]{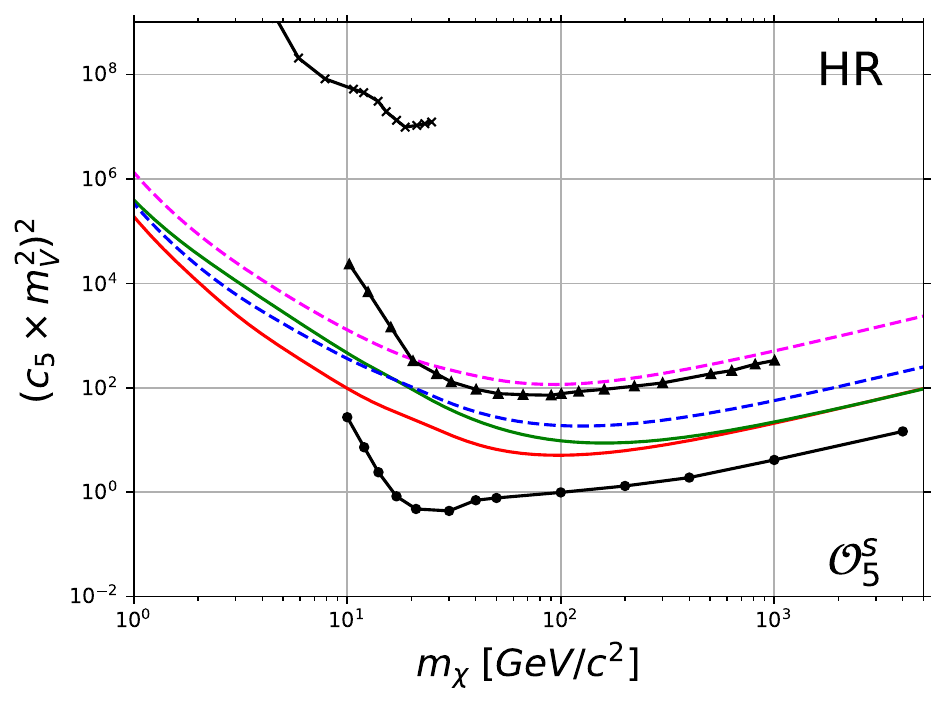}
    \end{subfigure}
    \begin{subfigure}{0.32\textwidth}
    \includegraphics[width=\linewidth]{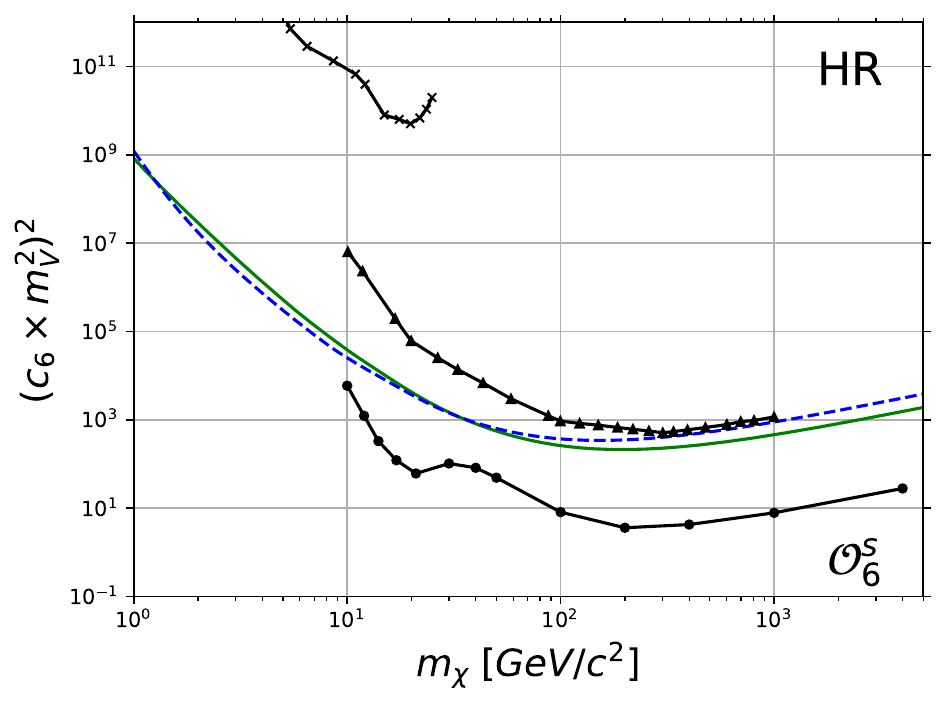}
    \end{subfigure}
    \begin{subfigure}{0.32\textwidth}
    \includegraphics[width=\linewidth]{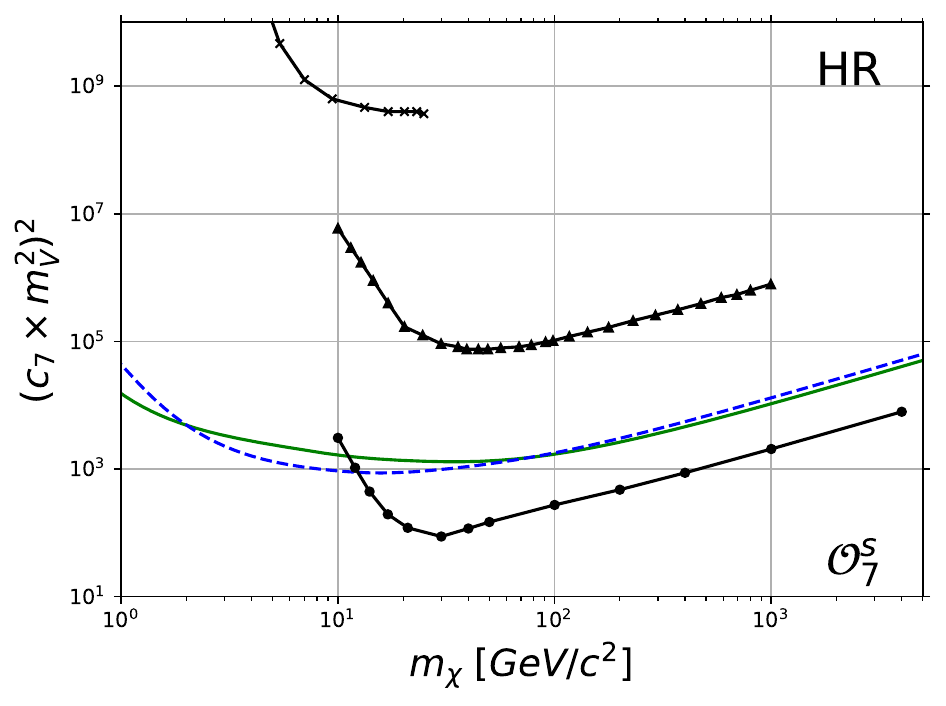}
    \end{subfigure}

    \vspace{0.1ex}

    \begin{subfigure}{0.32\textwidth}
    \includegraphics[width=\linewidth]{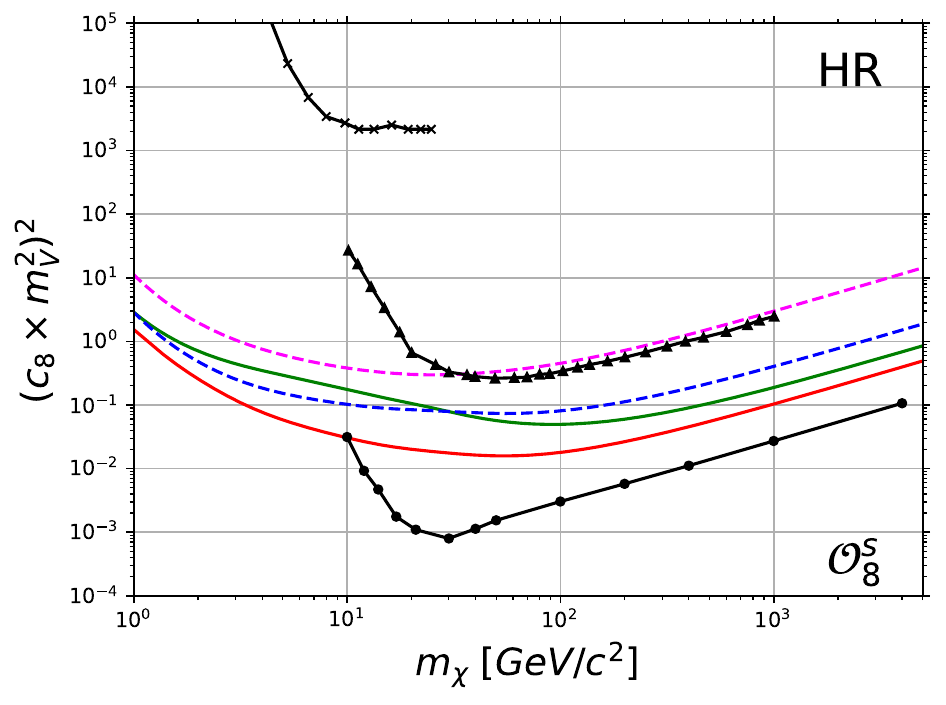}
    \end{subfigure}
    \begin{subfigure}{0.32\textwidth}
    \includegraphics[width=\linewidth]{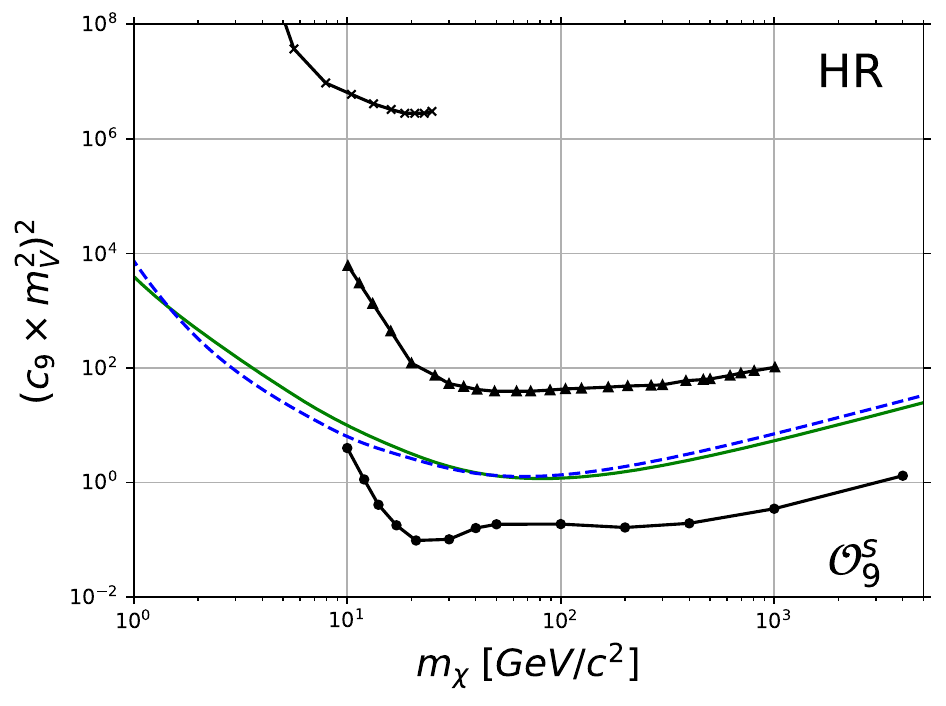}
    \end{subfigure}
    \begin{subfigure}{0.32\textwidth}
    \includegraphics[width=\linewidth]{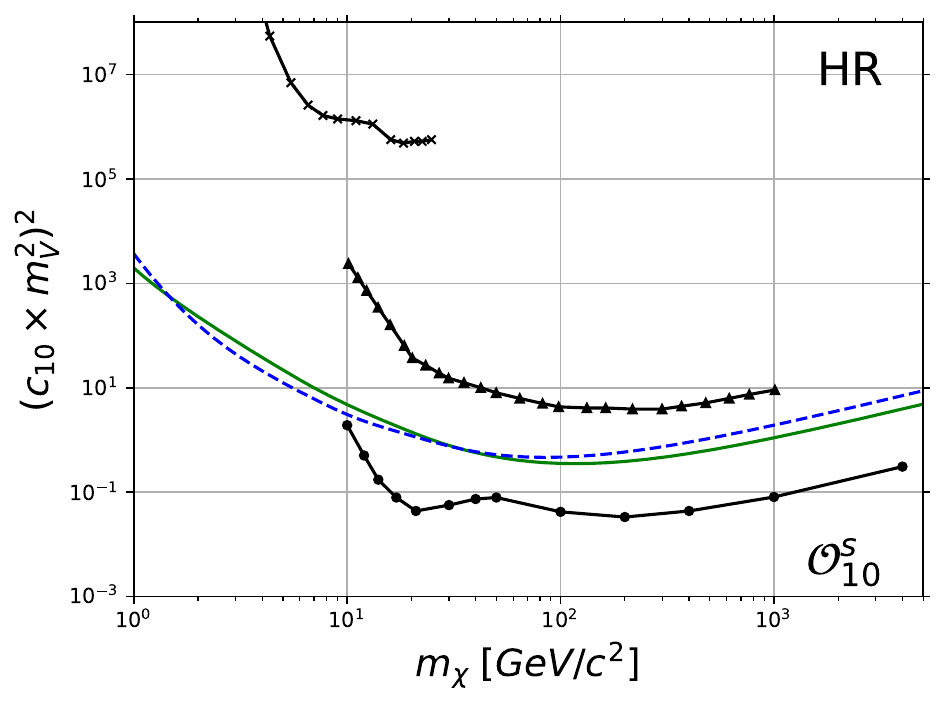}
    \end{subfigure}

    \vspace{0.1ex}

    \begin{subfigure}{0.32\textwidth}
    \includegraphics[width=\linewidth]{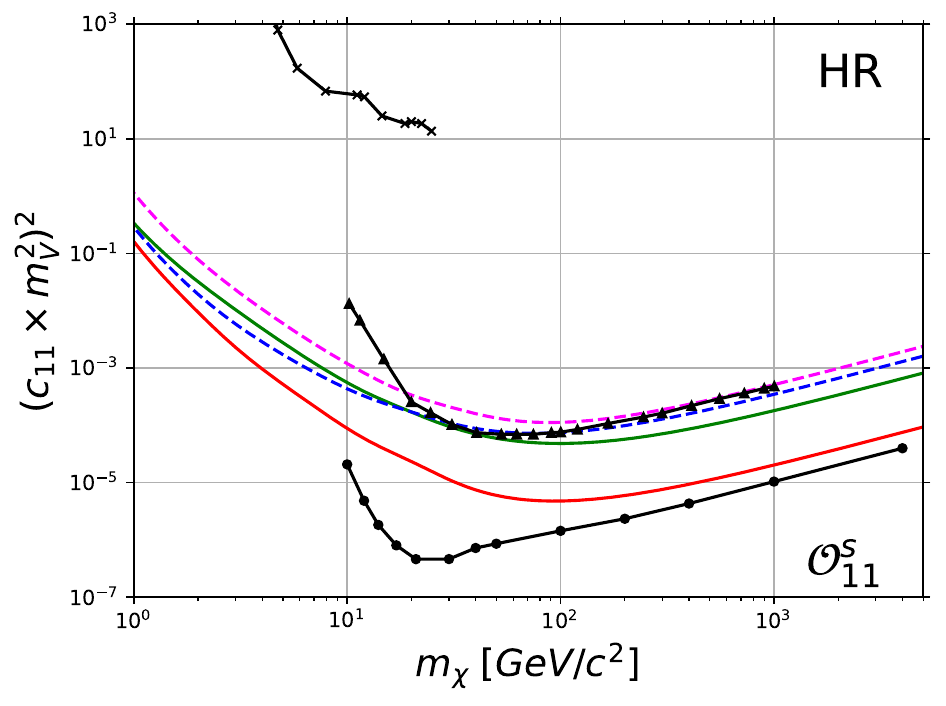}
    \end{subfigure}
    \begin{subfigure}{0.32\textwidth}
    \includegraphics[width=\linewidth]{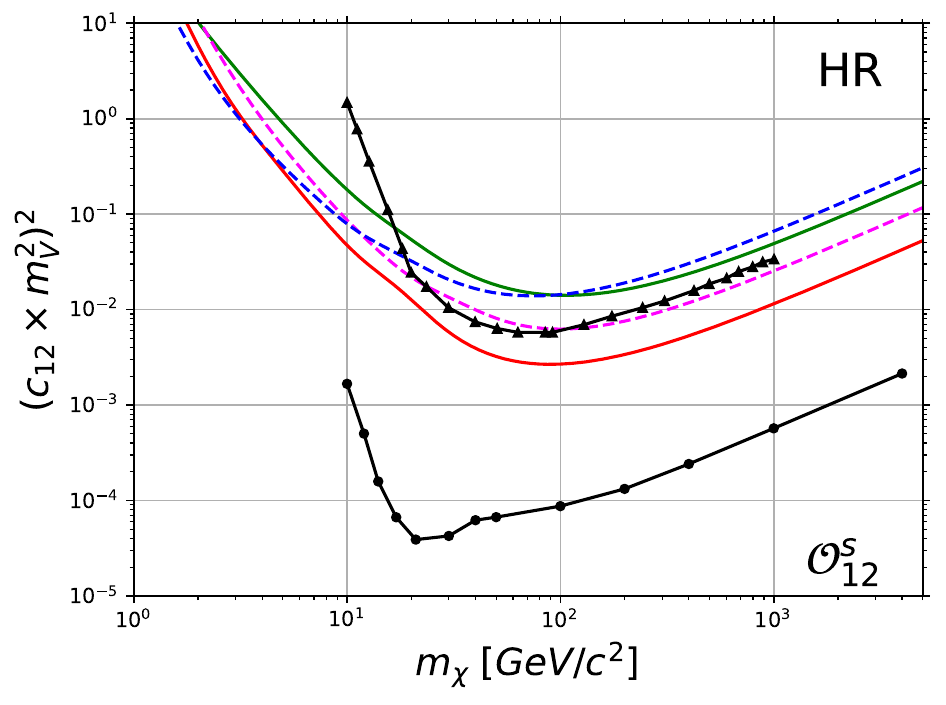}
    \end{subfigure}
    \begin{subfigure}{0.32\textwidth}
    \includegraphics[width=\linewidth]{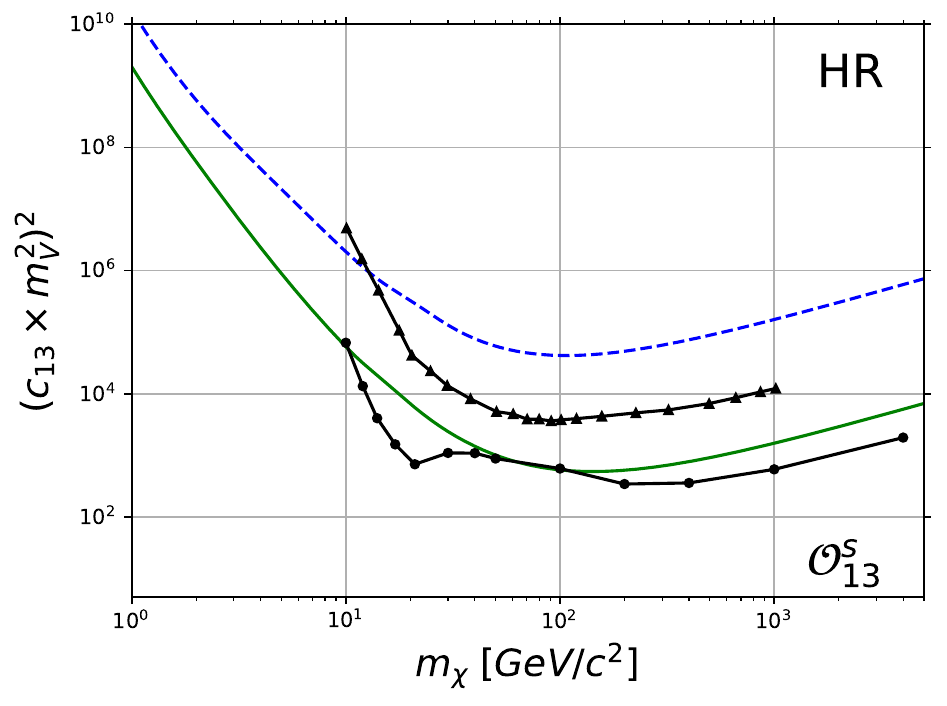}
    \end{subfigure}

    \vspace{0.1ex}

    \begin{subfigure}{0.32\textwidth}
    \includegraphics[width=\linewidth]{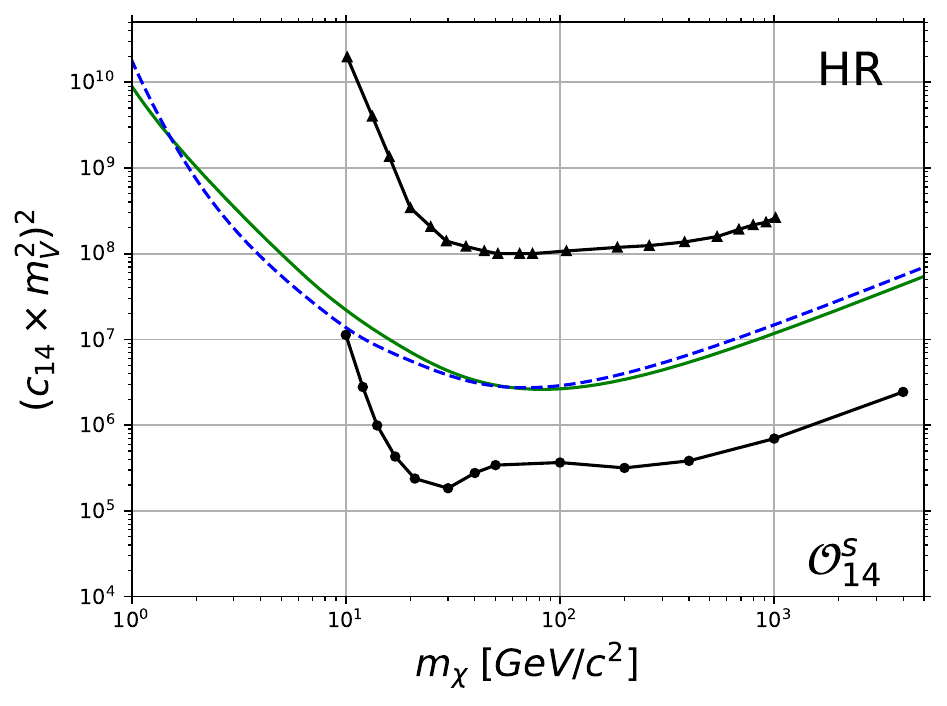}
    \end{subfigure}
    \begin{subfigure}{0.32\textwidth}
    \includegraphics[width=\linewidth]{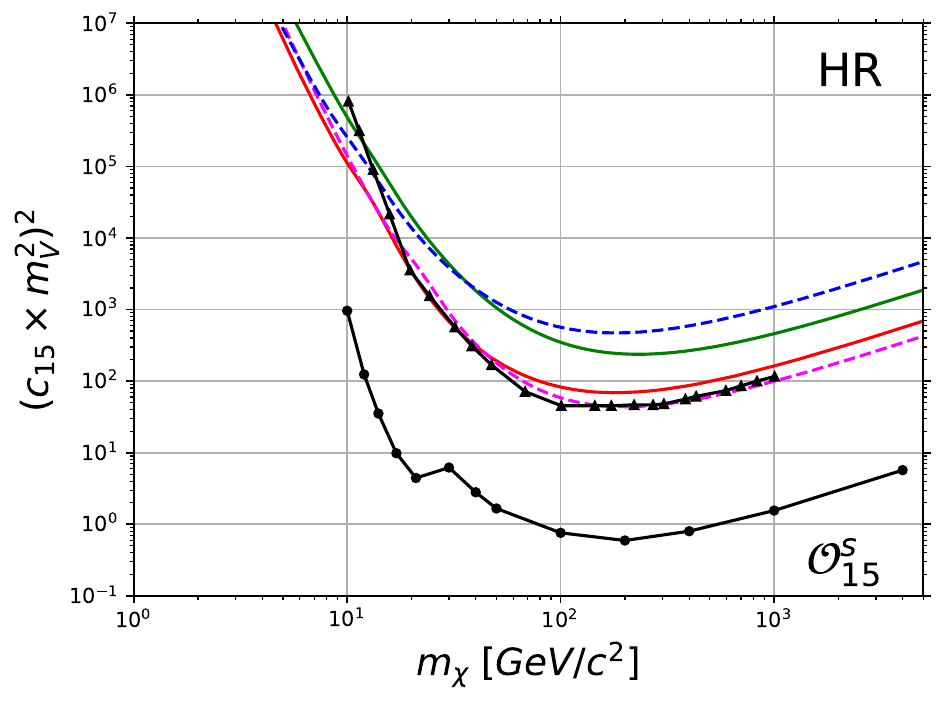}
    \end{subfigure}\hspace{4.55em}
    \begin{subfigure}{0.21\textwidth}
    \includegraphics[width=\linewidth]{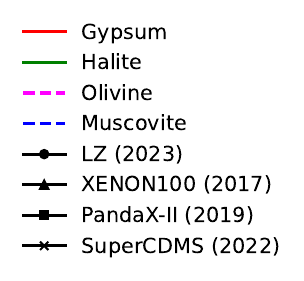} 
    \end{subfigure}
    \caption{Projected 90\% confidence level upper limits on the dimensionless isoscalar WIMP--nucleon NREFT coupling constants for elastic scattering, in the HR scenario. The solid (dashed) lines indicate minerals with $C^{238}=10^{-11}$ g/g ($C^{238}=10^{-10}$ g/g), see Table \ref{tab:U238_concentration}. Black lines show the NREFT results from conventional DD experiments: the 90\% confidence level upper limits from XENON100 \cite{XENON:2017fdd}, LUX--ZEPLIN \cite{LZ:2023lvz}, PandaX--II \cite{PandaX-II:2018woa}, as well as the 95\% Bayesian credible region of the two-dimensional marginalized posterior distribution from SuperCDMS \cite{SuperCDMS:2022crd}. Figures adapted from Ref.~{\cite{Theodosopoulos:2026ehn}}.}
    \label{fig:HR}
\end{figure}

\begin{figure}
    \captionsetup{justification=raggedright,singlelinecheck=false}
    \centering
    \begin{subfigure}{0.32\textwidth}
    \includegraphics[width=\linewidth]{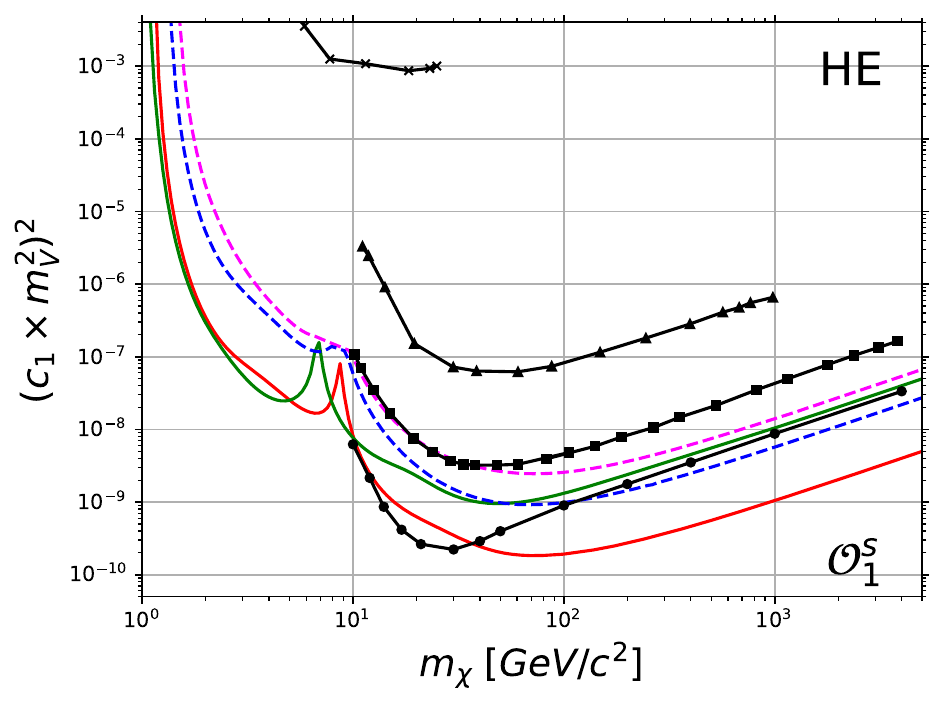}
    \end{subfigure}
    \begin{subfigure}{0.32\textwidth}
    \includegraphics[width=\linewidth]{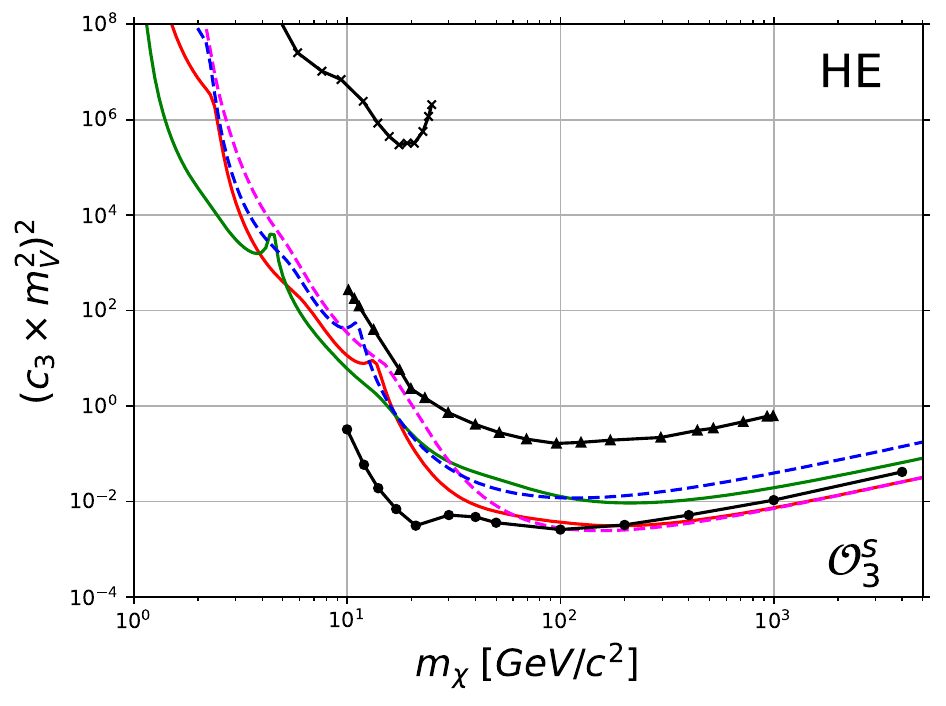}
    \end{subfigure}
    \begin{subfigure}{0.32\textwidth}
    \includegraphics[width=\linewidth]{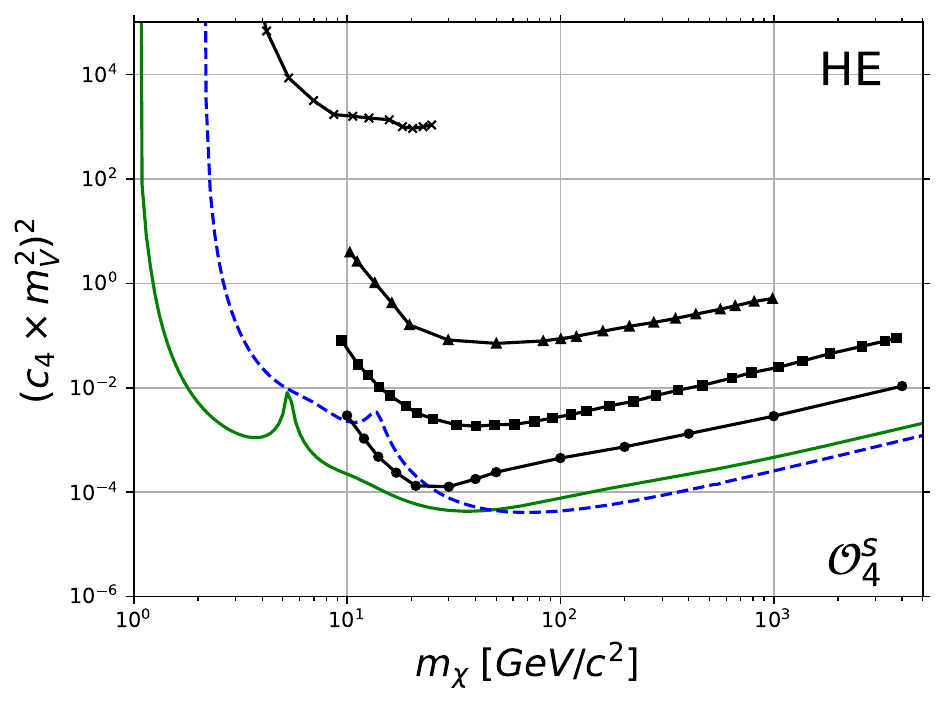}
    \end{subfigure}

    \vspace{0.1ex}

    \begin{subfigure}{0.32\textwidth}
    \includegraphics[width=\linewidth]{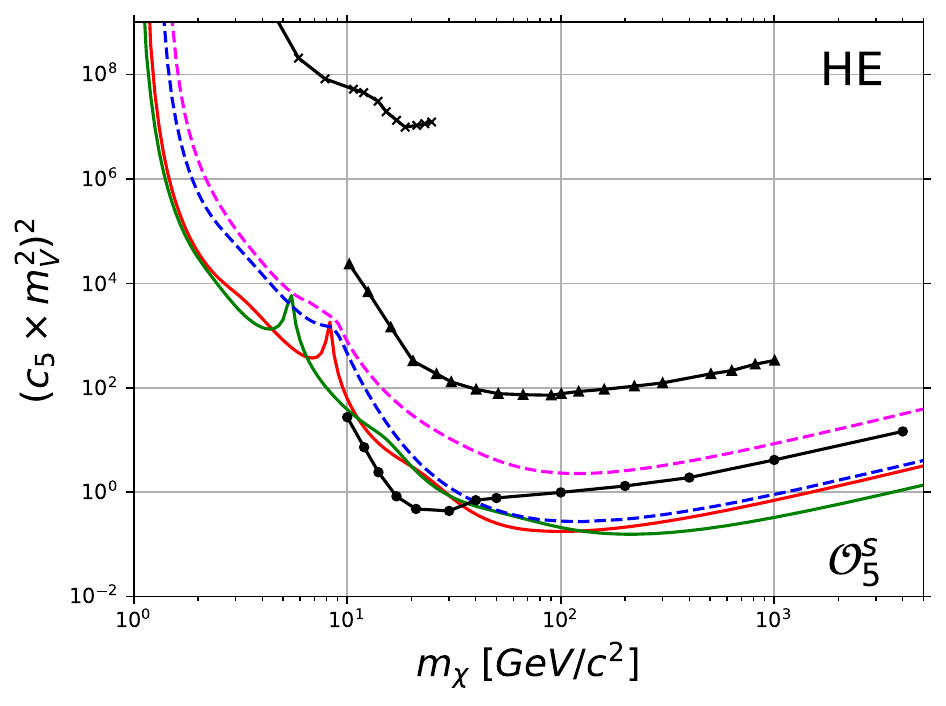}
    \end{subfigure}
    \begin{subfigure}{0.32\textwidth}
    \includegraphics[width=\linewidth]{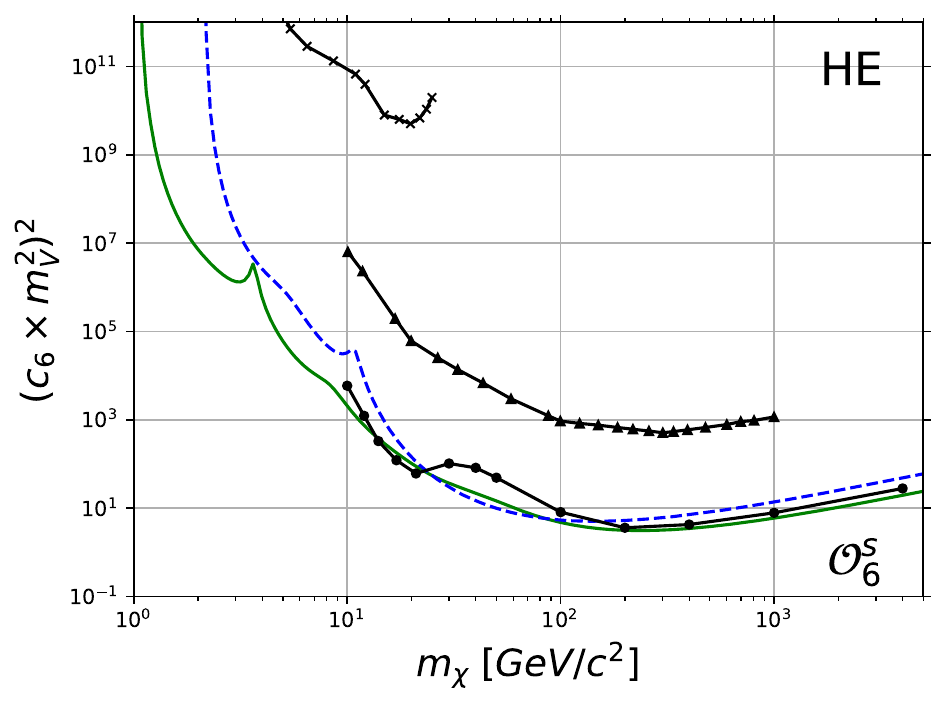}
    \end{subfigure}
    \begin{subfigure}{0.32\textwidth}
    \includegraphics[width=\linewidth]{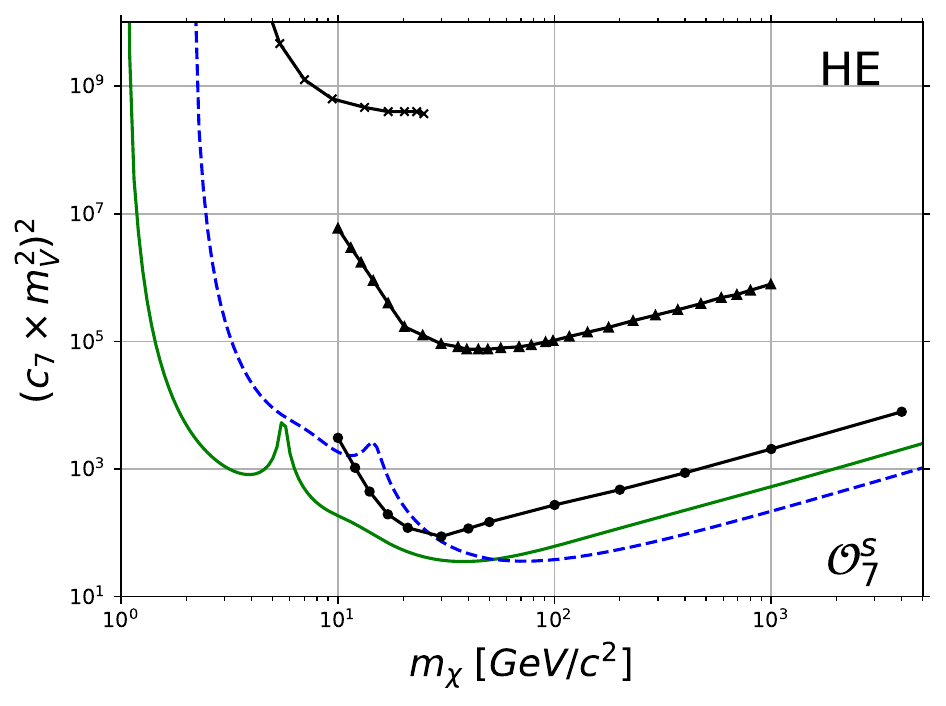}
    \end{subfigure}

    \vspace{0.1ex}

    \begin{subfigure}{0.32\textwidth}
    \includegraphics[width=\linewidth]{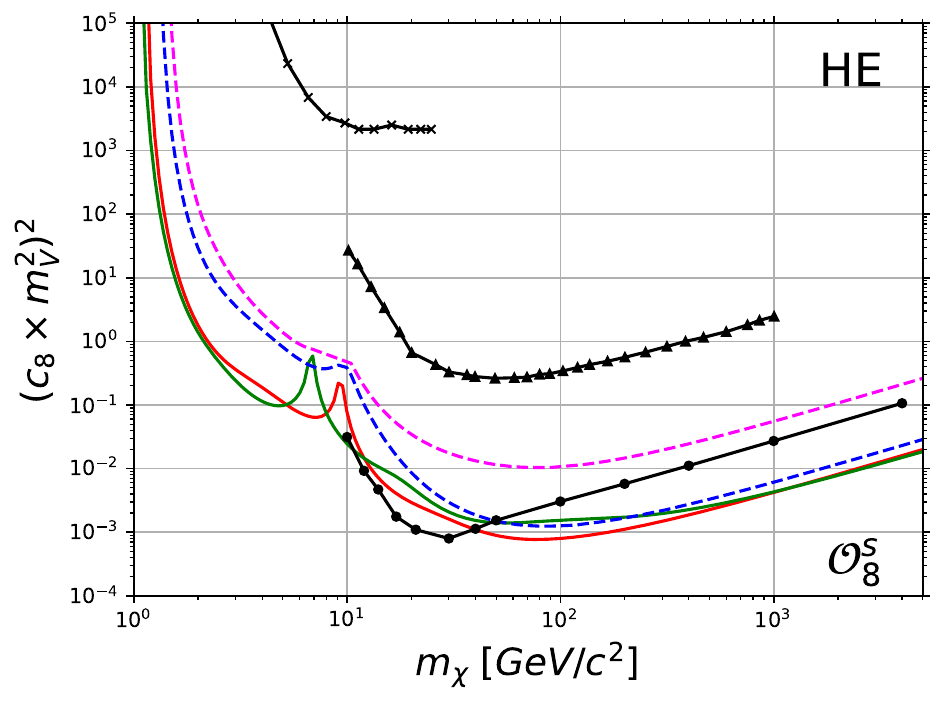}
    \end{subfigure}
    \begin{subfigure}{0.32\textwidth}
    \includegraphics[width=\linewidth]{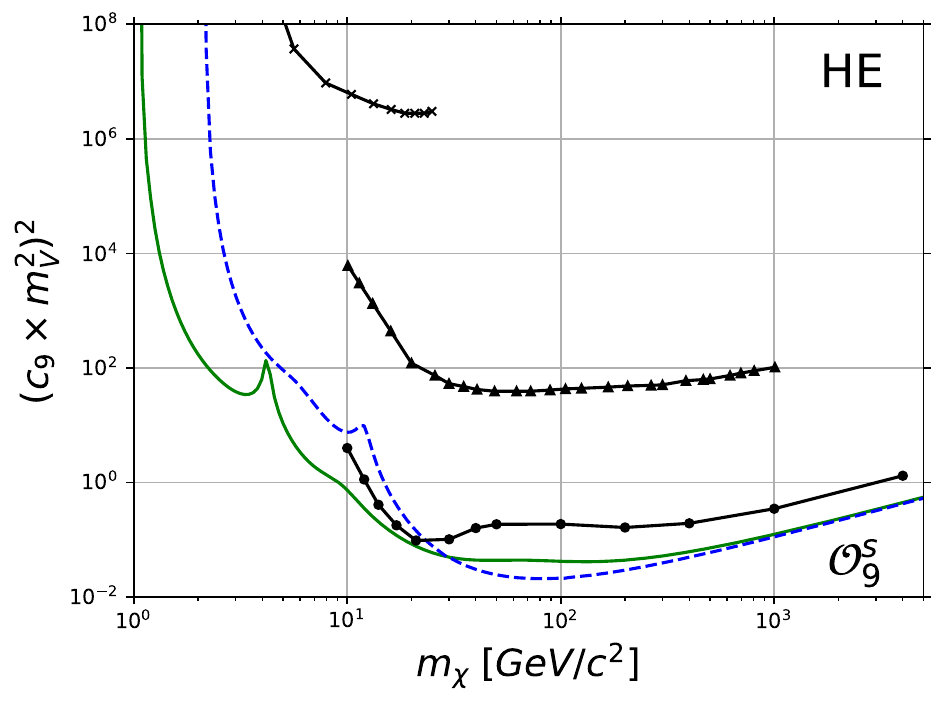}
    \end{subfigure}
    \begin{subfigure}{0.32\textwidth}
    \includegraphics[width=\linewidth]{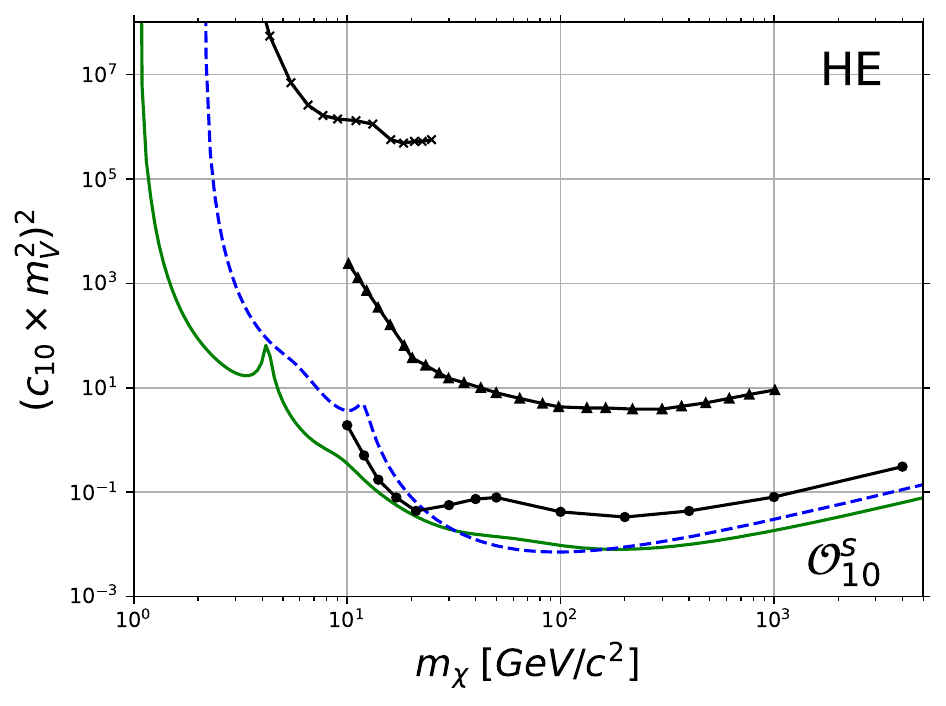}
    \end{subfigure}

    \vspace{0.1ex}

    \begin{subfigure}{0.32\textwidth}
    \includegraphics[width=\linewidth]{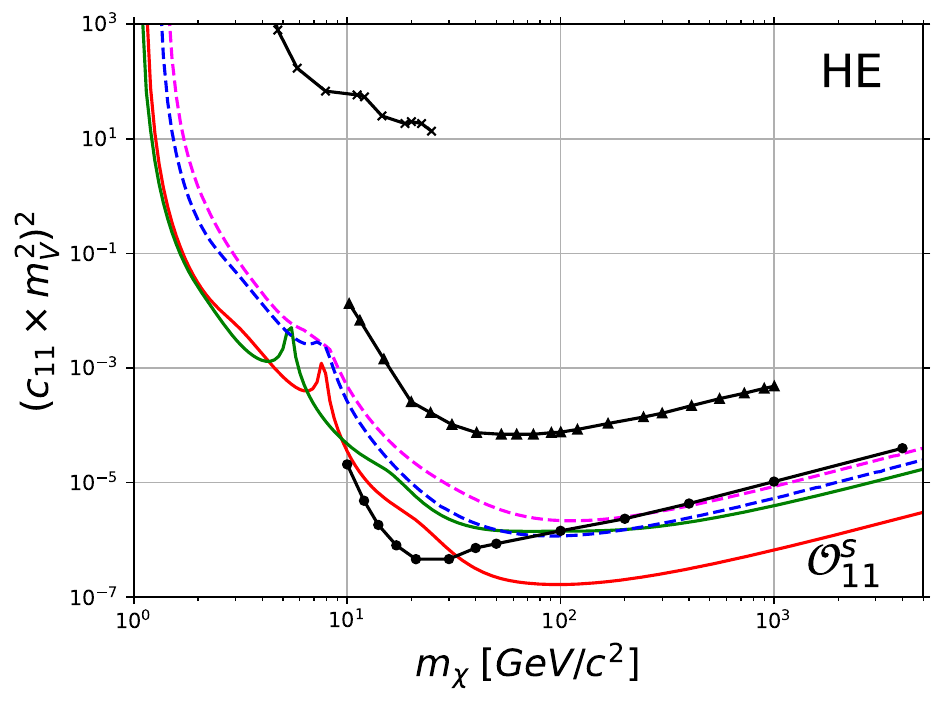}
    \end{subfigure}
    \begin{subfigure}{0.32\textwidth}
    \includegraphics[width=\linewidth]{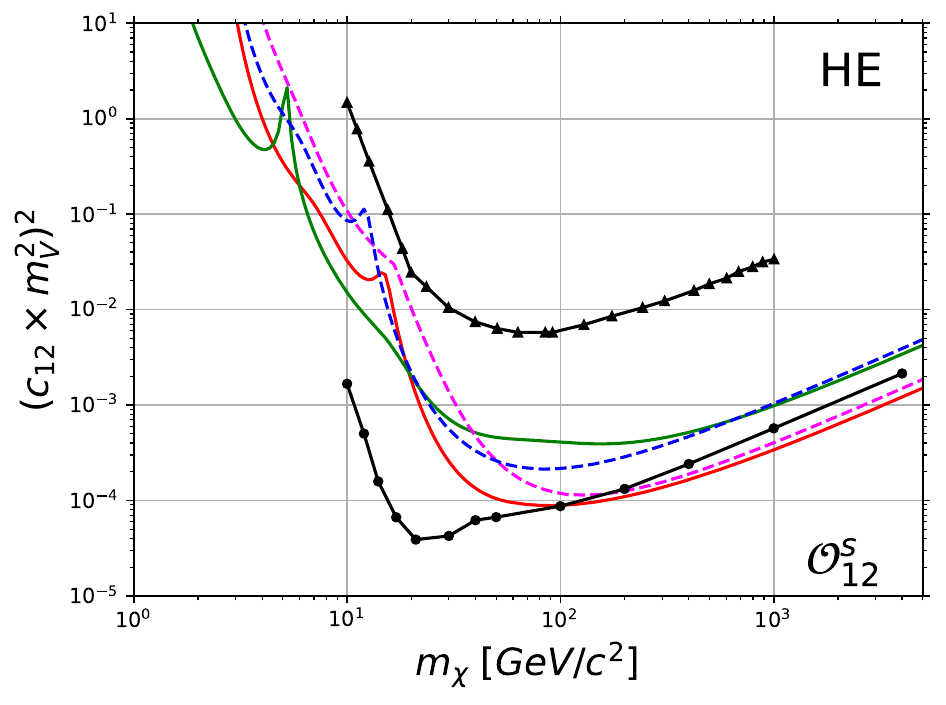}
    \end{subfigure}
    \begin{subfigure}{0.32\textwidth}
    \includegraphics[width=\linewidth]{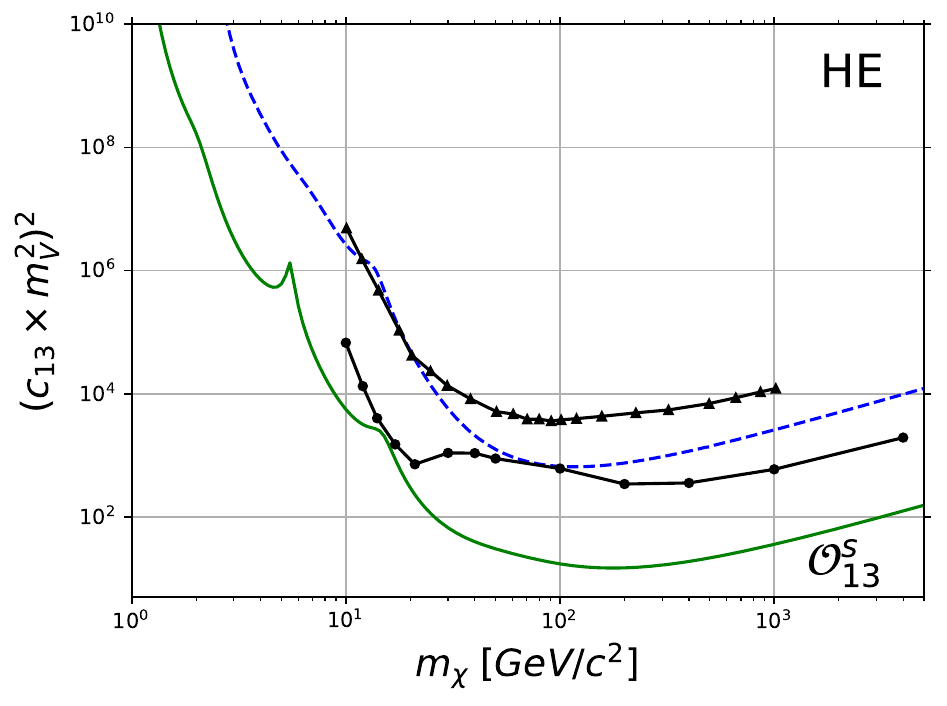}
    \end{subfigure}

    \vspace{0.1ex}

    \begin{subfigure}{0.32\textwidth}
    \includegraphics[width=\linewidth]{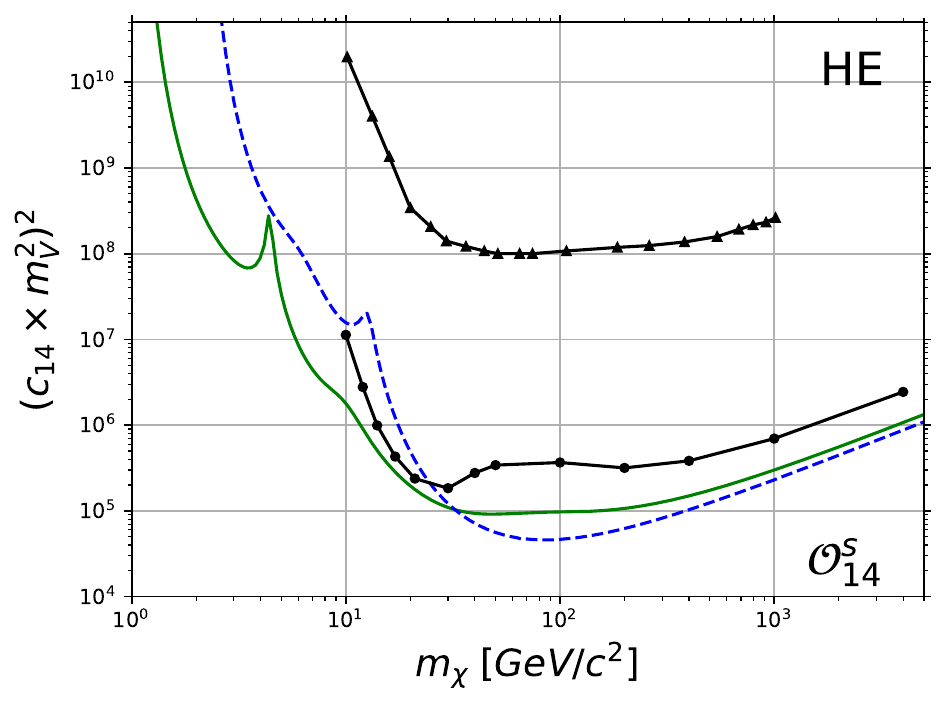}
    \end{subfigure}
    \begin{subfigure}{0.32\textwidth}
    \includegraphics[width=\linewidth]{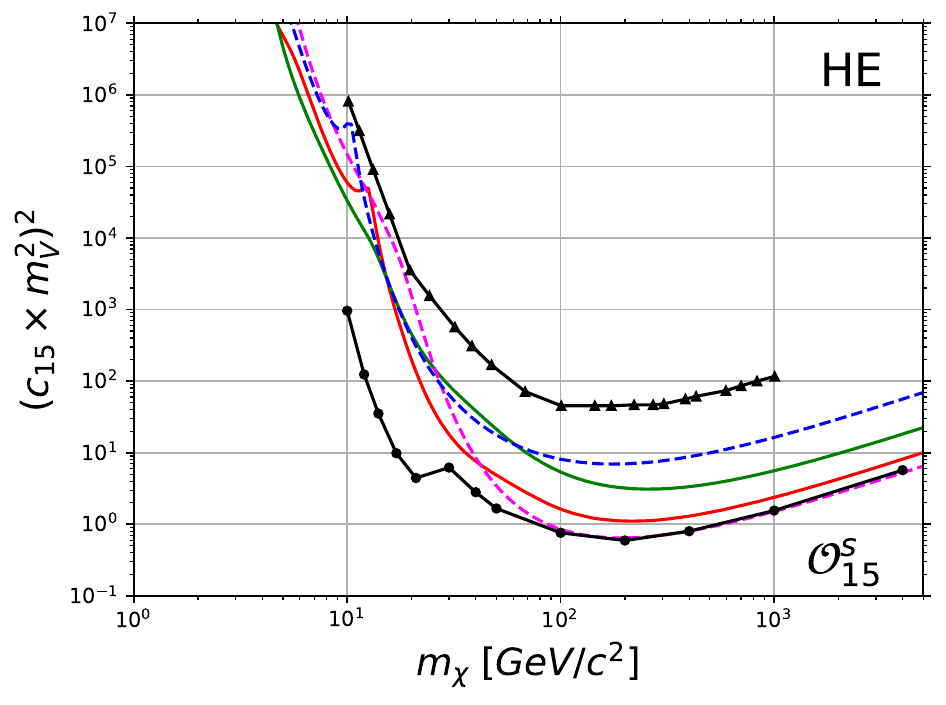}
    \end{subfigure}\hspace{4.55em}
    \begin{subfigure}{0.21\textwidth}
    \includegraphics[width=\linewidth]{figures/legend.pdf} 
    \end{subfigure}
    \caption{Projected 90\% confidence level upper limits on the dimensionless isoscalar WIMP--nucleon NREFT coupling constants for elastic scattering, in the HE scenario. The solid (dashed) lines indicate minerals with $C^{238}=10^{-11}$ g/g ($C^{238}=10^{-10}$ g/g), see Table \ref{tab:U238_concentration}. Black lines present the NREFT results from conventional DD experiments: the 90\% confidence level upper limits from XENON100 \cite{XENON:2017fdd}, LUX--ZEPLIN \cite{LZ:2023lvz}, PandaX--II \cite{PandaX-II:2018woa}, as well as the 95\% Bayesian credible region of the two-dimensional marginalized posterior distribution from SuperCDMS \cite{SuperCDMS:2022crd}. Figures adapted from Ref.~{\cite{Theodosopoulos:2026ehn}}.}
    \label{fig:HE}
\end{figure}

Figs.~\ref{fig:HR} and \ref{fig:HE} show the projected $90\%$ confidence level exclusion limits on the isoscalar WIMP--nucleon NREFT coupling constants ($c^{p}=c^{n}$) for elastic scattering, as a function of the DM mass in the range
$1$--$5000~\mathrm{GeV}/c^{2}$. The results are shown for the HR scenario ($M=10~\mathrm{mg}$, $\sigma_{x}=1~\mathrm{nm}$) and the HE scenario ($M=100~\mathrm{g}$, $\sigma_{x}=15~\mathrm{nm}$), respectively, and are taken from Ref.~\cite{Theodosopoulos:2026ehn}. The coupling constants are expressed in weak-interaction units, such that $c_j\,m_{v}^{2}$ is dimensionless, with $m_{v}=246.2~\mathrm{GeV}$ denoting the electroweak scale. We consider four target minerals commonly found in nature (see Table~\ref{tab:U238_concentration}): two ultra-basic rocks (UBRs), olivine and muscovite, and two marine evaporites (MEs), halite and gypsum. Muscovite and gypsum contain hydrogen, while olivine and halite do not.

As discussed in the previous sections, the HR scenario provides superior sensitivity for low DM masses, $m_{\chi}\lesssim10~\mathrm{GeV}/c^{2}$, where solar neutrinos dominate the background. In this regime, the excellent track--length resolution of the HR scenario is essential to discriminate between signal and background spectral shapes. Consequently, for light DM particles, the projected limits on the NREFT coupling constants in the HR scenario (Fig.~\ref{fig:HR}) are significantly stronger than those obtained in the HE scenario (Fig.~\ref{fig:HE}), where the limited resolution renders the couplings effectively unconstrained. For larger DM masses, $m_{\chi}\gtrsim10~\mathrm{GeV}/c^{2}$, radiogenic backgrounds become dominant and sensitivity is driven primarily by exposure. In this regime, the HE scenario yields stronger constraints.

Comparing different target materials, hydrogen-bearing minerals generally provide improved sensitivity due to the efficient moderation of radiogenic neutrons by hydrogen. In addition, MEs typically outperform UBRs, since we assume a lower $^{238}\mathrm{U}$ concentration for MEs ($C^{238}=10^{-11}~\mathrm{g/g}$) than for UBRs ($C^{238}=10^{-10}~\mathrm{g/g}$). This effect is illustrated by comparing gypsum and olivine. As discussed in Ref.~\cite{Theodosopoulos:2026ehn}, both gypsum and olivine consist of isotopes with spin-zero ground states and therefore produce non-vanishing spectra only for a subset of NREFT operators, namely $\mathcal{O}_{1}^{s}$, $\mathcal{O}_{3}^{s}$, $\mathcal{O}_{5}^{s}$, $\mathcal{O}_{8}^{s}$, $\mathcal{O}_{11}^{s}$, $\mathcal{O}_{12}^{s}$, and $\mathcal{O}_{15}^{s}$. Although their mass numbers are comparable and the corresponding signal spectra are similar, gypsum yields stronger sensitivity in both readout scenarios due to its hydrogen content and lower uranium concentration, which together suppress radiogenic backgrounds. Halite and muscovite contain isotopes with nonzero-spin ground states and therefore exhibit nonzero spectra for all NREFT operators.

Figures~\ref{fig:HR} and \ref{fig:HE} also include existing limits from conventional direct-detection experiments, namely XENON100~\cite{XENON:2017fdd}, LUX--ZEPLIN~\cite{LZ:2023lvz}, and PandaX--II~\cite{PandaX-II:2018woa}, as well as the $95\%$ Bayesian credible regions reported by SuperCDMS~\cite{SuperCDMS:2022crd}. The latter are comparable to frequentist $90\%$ confidence level limits, as discussed in Ref.~\cite{SuperCDMS:2022crd}. Paleo-detectors with high readout resolution can probe WIMP--nucleon interactions well below current experimental bounds for $m_{\chi}\lesssim10~\mathrm{GeV}/c^{2}$. At higher masses, $m_{\chi}\gtrsim10~\mathrm{GeV}/c^{2}$, the large exposures achievable in the HE scenario allow paleo-detectors to surpass the sensitivity of existing experiments. 

\begin{figure}
    \captionsetup{justification=raggedright,singlelinecheck=false}
    \centering
    \begin{subfigure}{0.32\textwidth}
    \includegraphics[width=\linewidth]{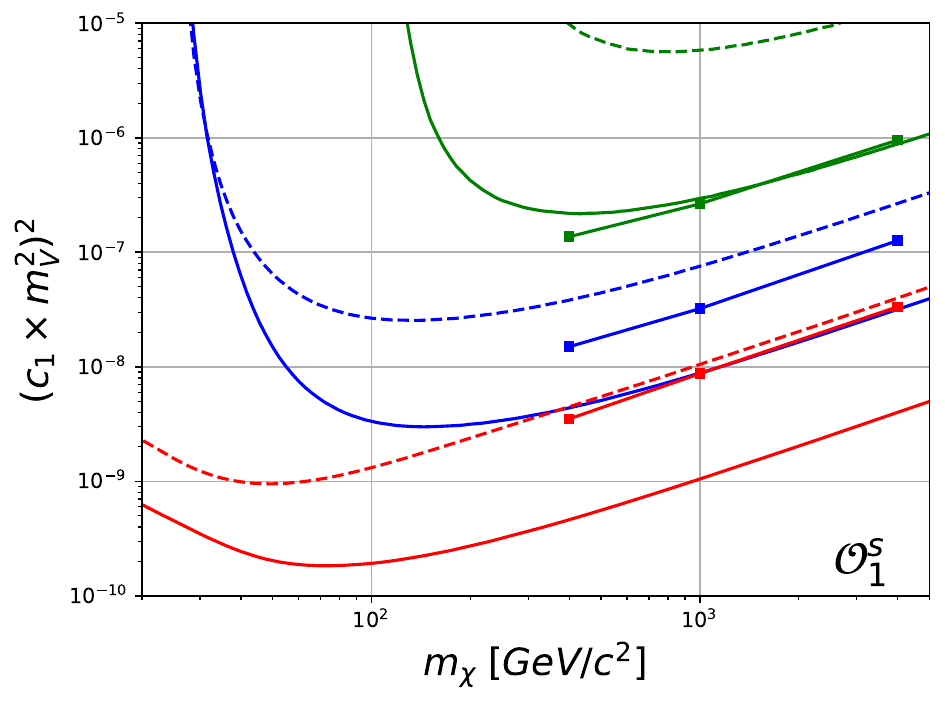}
    \end{subfigure}
    \begin{subfigure}{0.32\textwidth}
    \includegraphics[width=\linewidth]{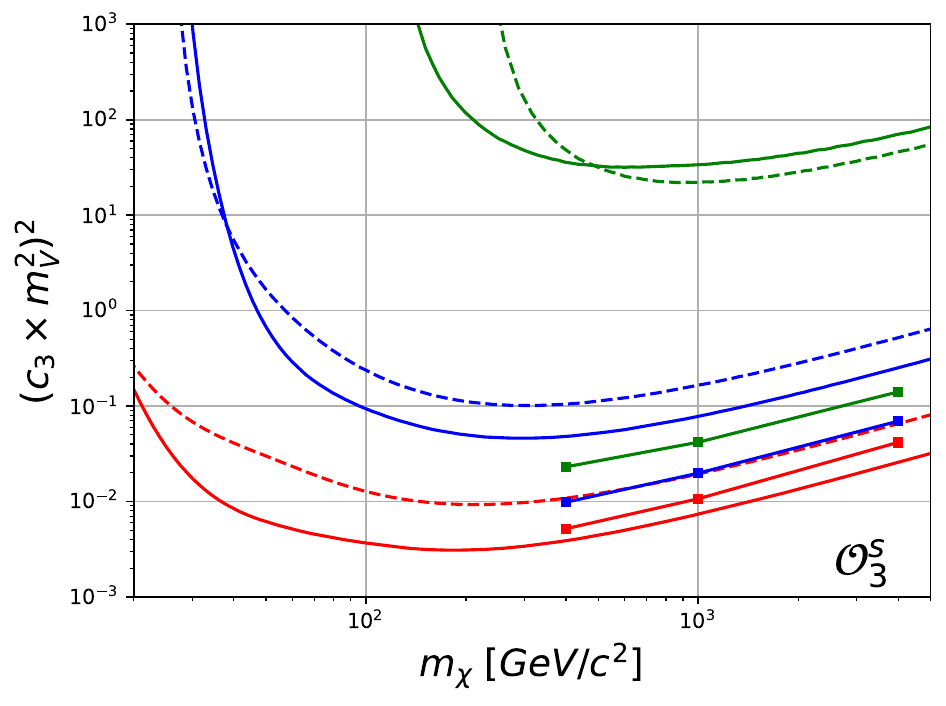}
    \end{subfigure}
    \begin{subfigure}{0.32\textwidth}
    \includegraphics[width=\linewidth]{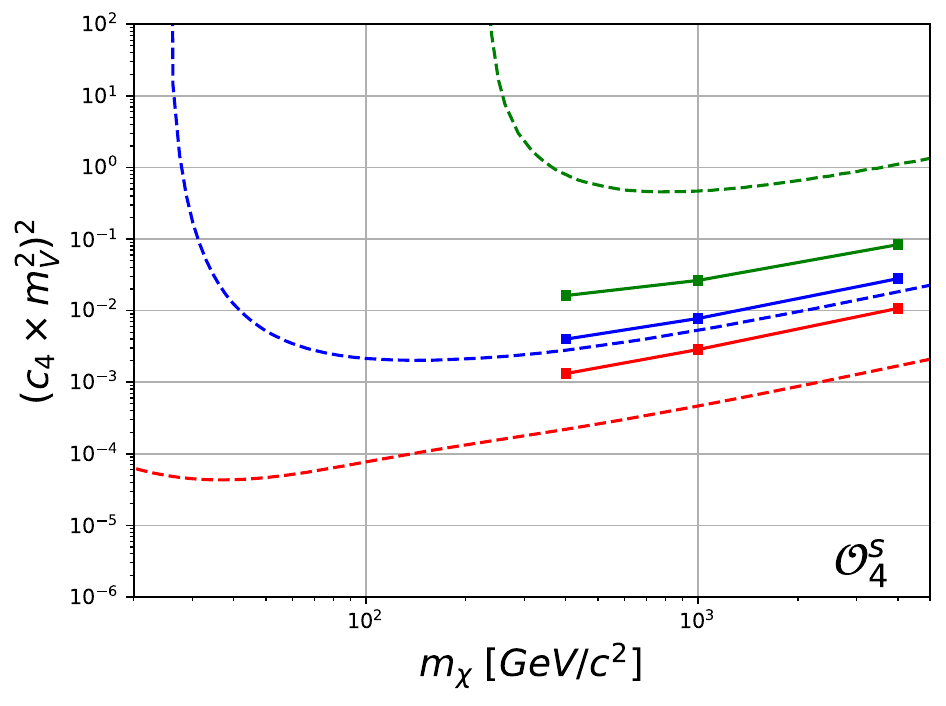}
    \end{subfigure}

    \vspace{0.1ex}

    \begin{subfigure}{0.32\textwidth}
    \includegraphics[width=\linewidth]{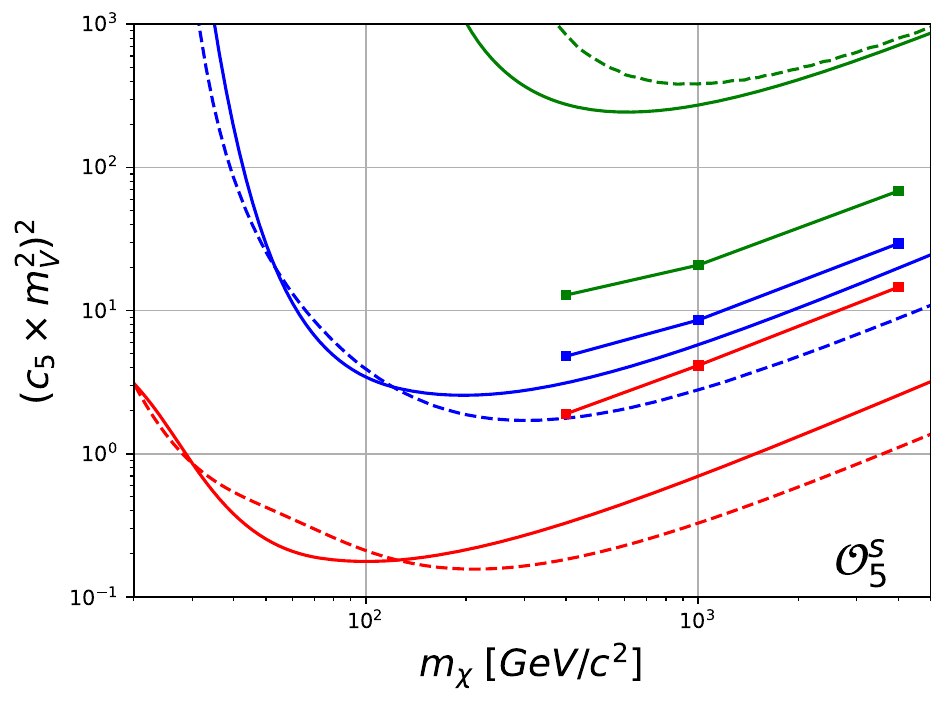}
    \end{subfigure}
    \begin{subfigure}{0.32\textwidth}
    \includegraphics[width=\linewidth]{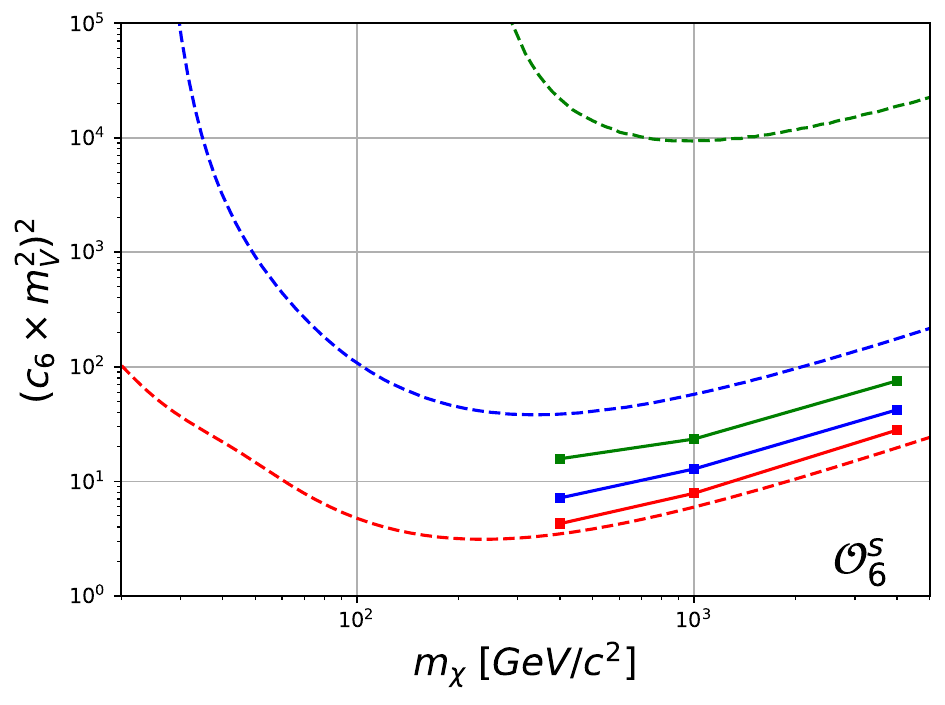}
    \end{subfigure}
    \begin{subfigure}{0.32\textwidth}
    \includegraphics[width=\linewidth]{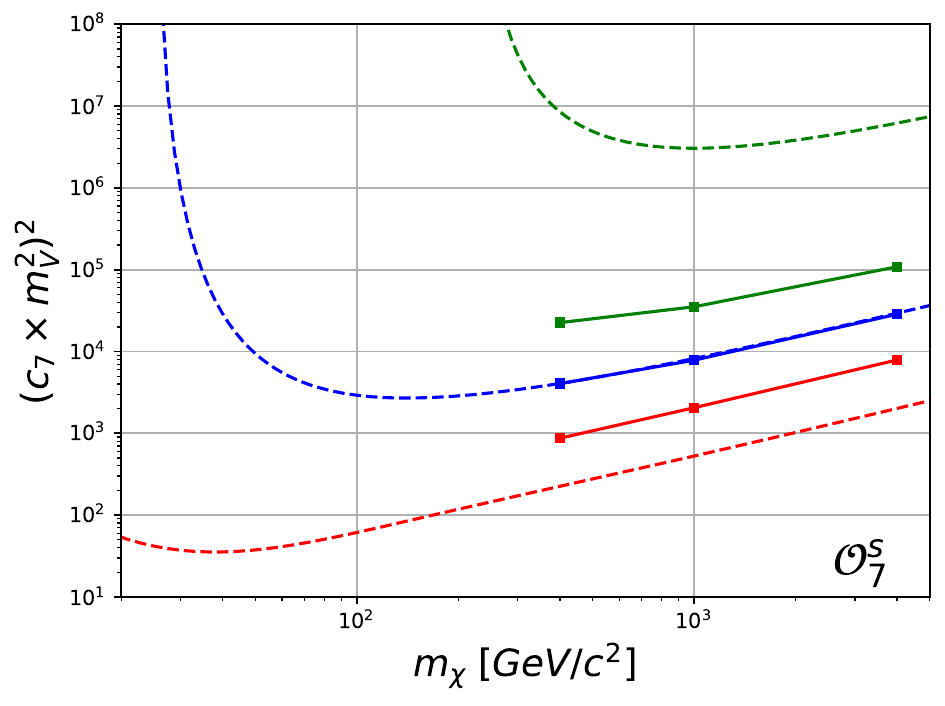}
    \end{subfigure}

    \vspace{0.1ex}

    \begin{subfigure}{0.32\textwidth}
    \includegraphics[width=\linewidth]{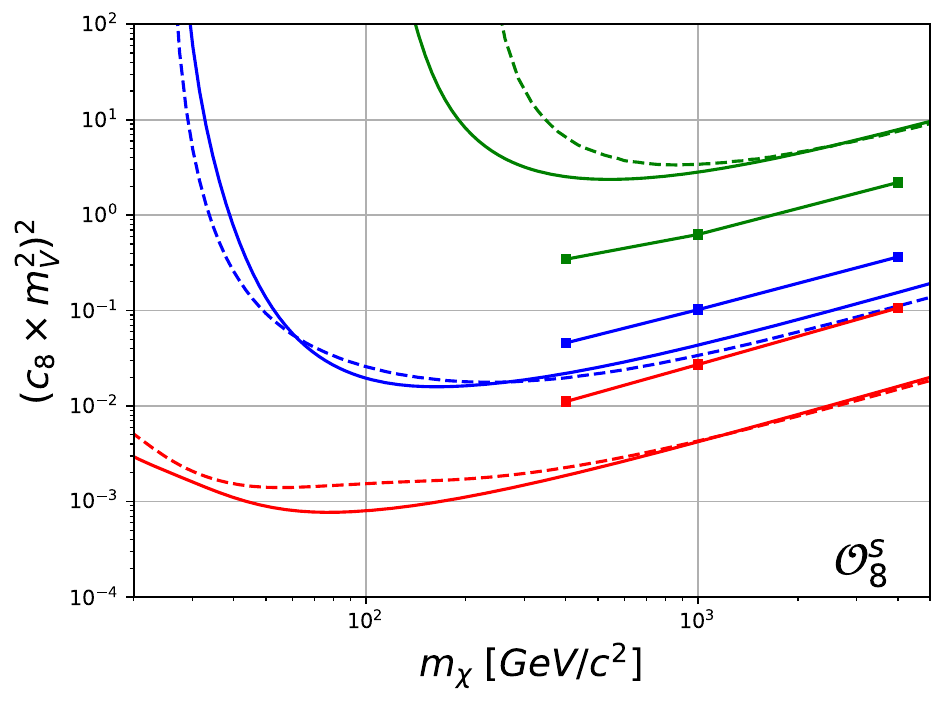}
    \end{subfigure}
    \begin{subfigure}{0.32\textwidth}
    \includegraphics[width=\linewidth]{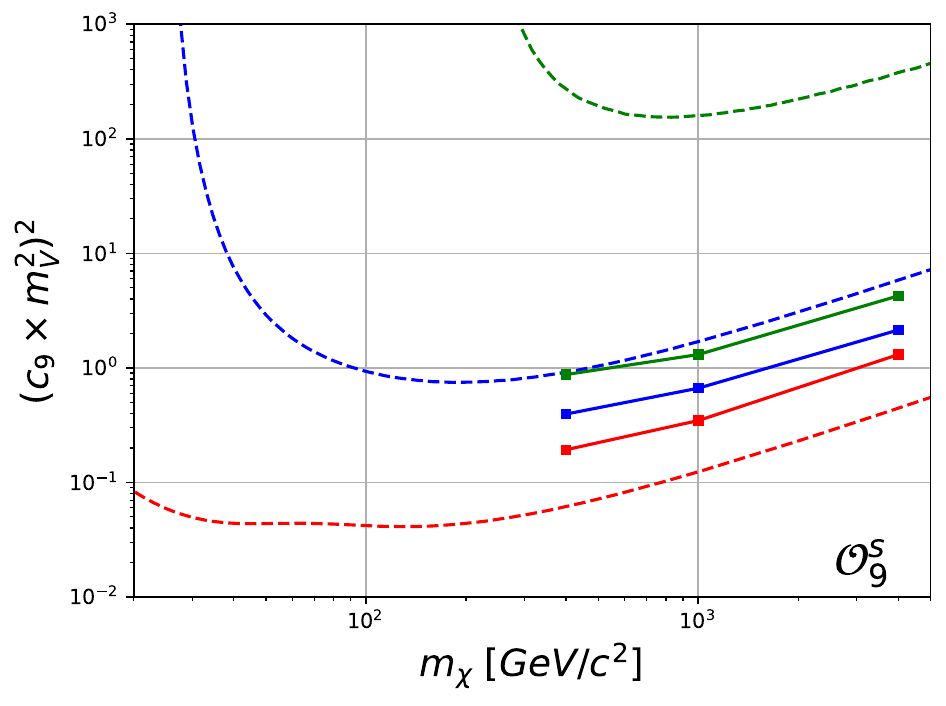}
    \end{subfigure}
    \begin{subfigure}{0.32\textwidth}
    \includegraphics[width=\linewidth]{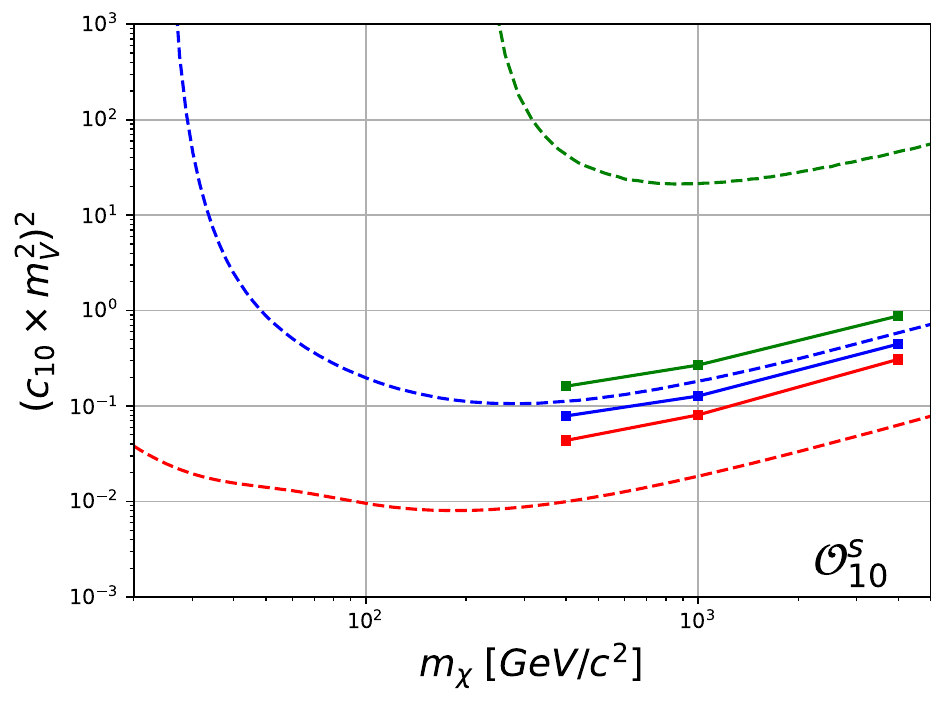}
    \end{subfigure}

    \vspace{0.1ex}

    \begin{subfigure}{0.32\textwidth}
    \includegraphics[width=\linewidth]{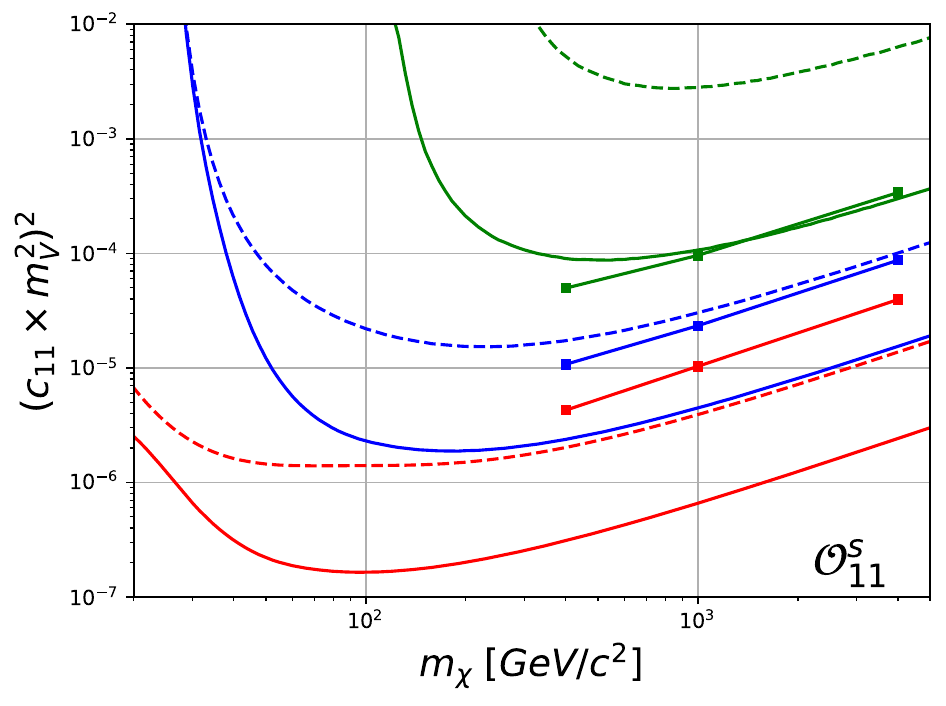}
    \end{subfigure}
    \begin{subfigure}{0.32\textwidth}
    \includegraphics[width=\linewidth]{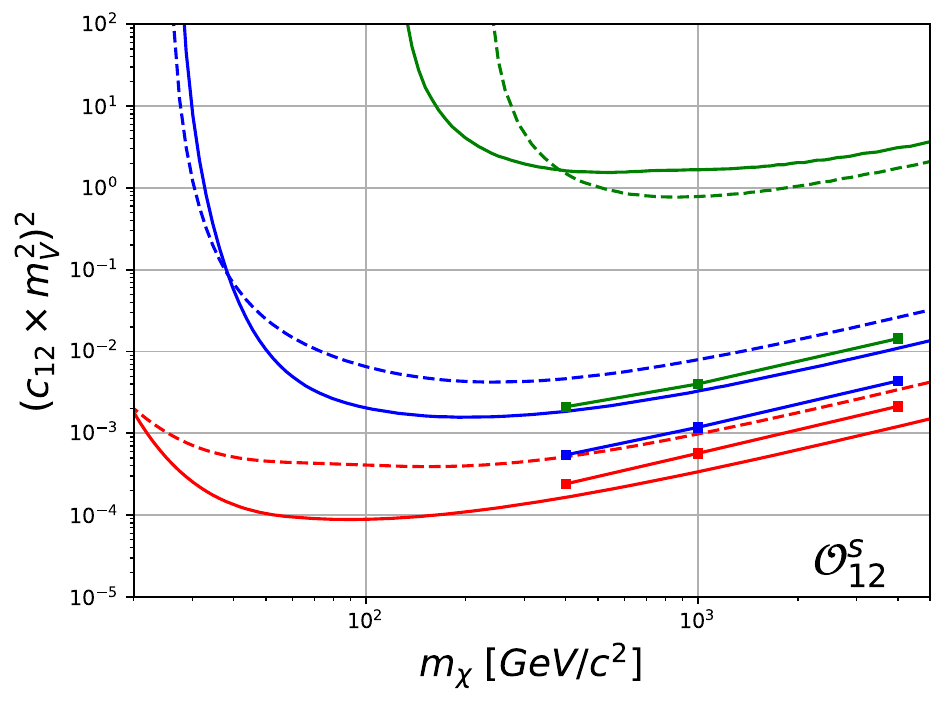}
    \end{subfigure}
    \begin{subfigure}{0.32\textwidth}
    \includegraphics[width=\linewidth]{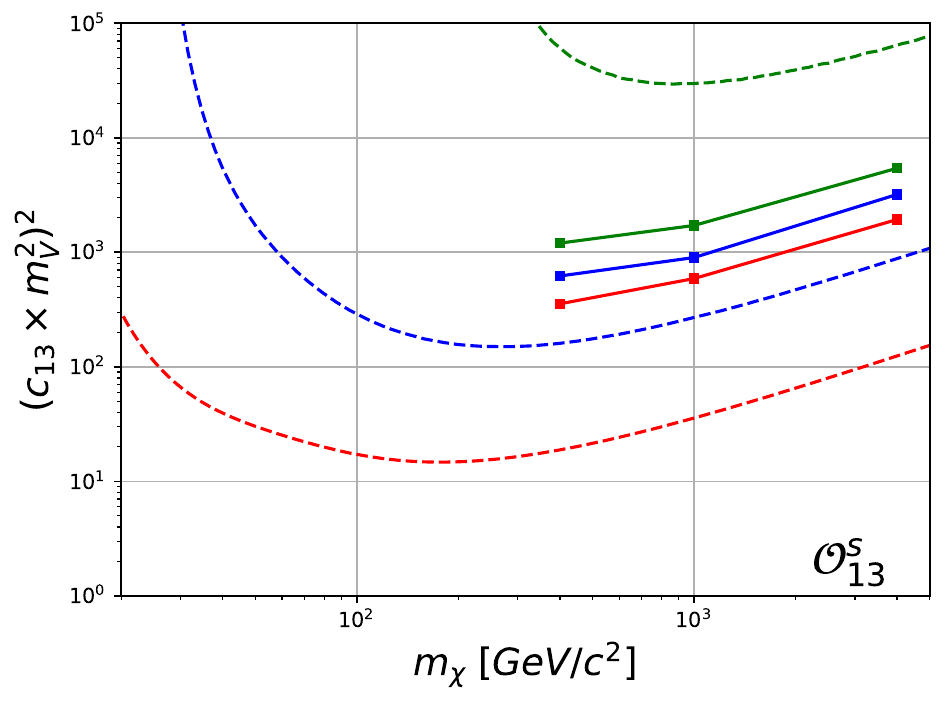}
    \end{subfigure}

    \vspace{0.1ex}

    \begin{subfigure}{0.32\textwidth}
    \includegraphics[width=\linewidth]{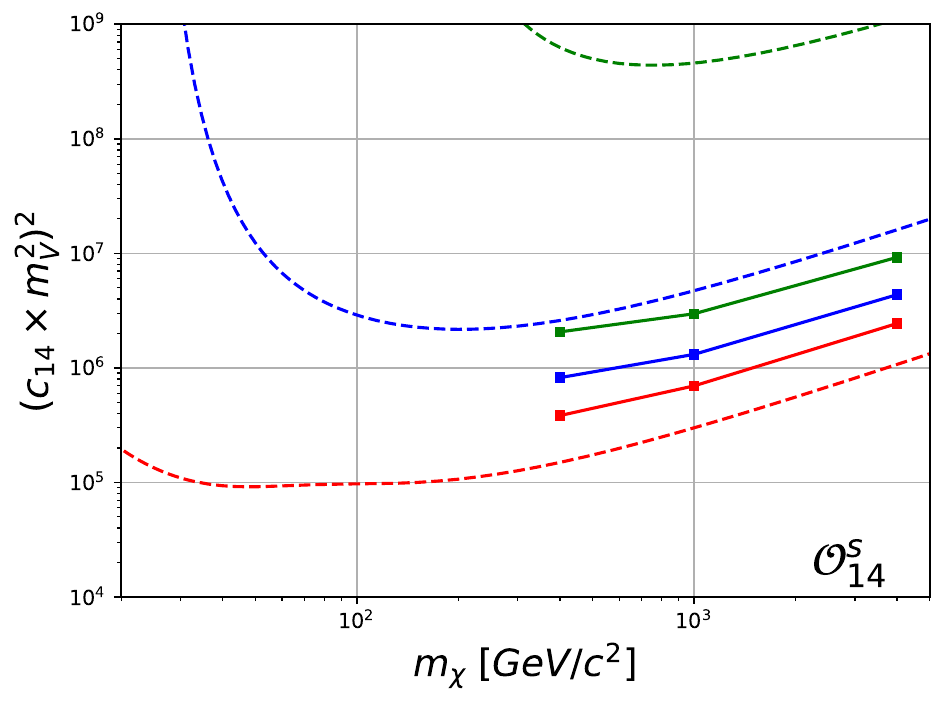}
    \end{subfigure}
    \begin{subfigure}{0.32\textwidth}
    \includegraphics[width=\linewidth]{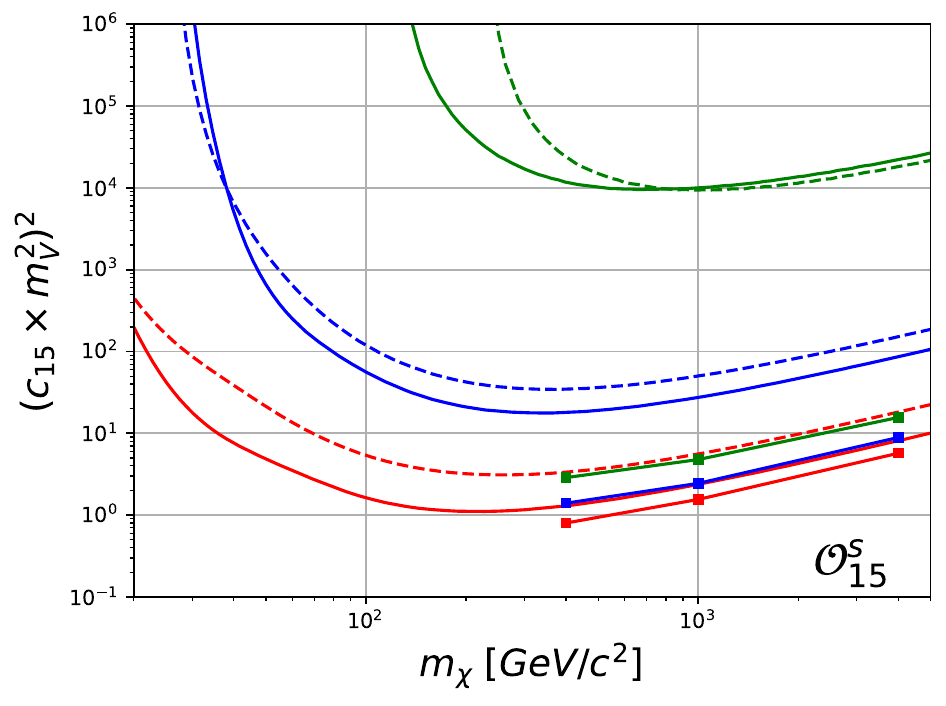}
    \end{subfigure}\hspace{4.55em}
    \begin{subfigure}{0.21\textwidth}
    \includegraphics[width=\linewidth]{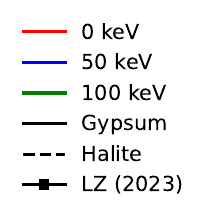} 
    \end{subfigure}
    \caption{Projected 90\% confidence–level upper limits on the dimensionless isoscalar WIMP–nucleon NREFT coupling constants for inelastic scattering, in the HE scenario. Colored lines correspond to different mass splittings $\delta_{m}$, while square marker show the NREFT limits from the LUX–ZEPLIN experiment \cite{LZ:2023lvz}. The target minerals are gypsum (solid lines) and halite (dashed lines). Figures adapted from Ref.~{\cite{Theodosopoulos:2026ehn}}.}
    \label{fig:inelastic}
\end{figure}

In addition to elastic interactions, we also investigate inelastic DM scattering. Fig.~\ref{fig:inelastic} shows the projected $90\%$ confidence level exclusion limits on the isoscalar WIMP--nucleon NREFT coupling constants ($c^{p}=c^{n}$) for inelastic scattering, as a function of the DM mass in the range
$20$--$5000~\mathrm{GeV}/c^{2}$. The results are presented for the HE scenario ($M=100~\mathrm{g}$, $\sigma_{x}=15~\mathrm{nm}$) and for mass splittings $\delta_{m}$ between $0$ and $100~\mathrm{keV}/c^{2}$, a range motivated by a broad class of WIMP models~\cite{Tucker-Smith:2001myb,Barello:2014uda}. The results are taken from Ref.~\cite{Theodosopoulos:2026ehn}.

We focus on gypsum and halite as representative target materials, whose dominant nuclei include Ca and Cl with masses of approximately $40~\mathrm{GeV}/c^{2}$ and $35~\mathrm{GeV}/c^{2}$, respectively (see Table~\ref{tab:U238_concentration}). For inelastic scattering with large mass splittings, the recoil track--length spectrum is strongly suppressed unless the DM particle is sufficiently heavy. Consequently, we restrict our study to the HE scenario, which favors sensitivity to higher DM masses. This behavior is evident in Fig.~\ref{fig:inelastic}: for large splittings, limits are obtained only at high $m_{\chi}$, while for $m_{\chi}\lesssim100~\mathrm{GeV}/c^{2}$ the coupling constants remain unconstrained when $\delta_{m}=100~\mathrm{keV}/c^{2}$.

Fig.~\ref{fig:inelastic} also includes constraints from the LUX--ZEPLIN experiment~\cite{LZ:2023lvz}. For spin-independent NREFT operators and a mass splitting of $\delta_{m}=50~\mathrm{keV}/c^{2}$, gypsum yields stronger limits than LUX--ZEPLIN, following trends similar to those observed in the elastic-scattering case. As shown in the top-left panel of Fig.~\ref{fig:Spectrum_binned_SI_inelastic}, the signal spectra in the HE scenario are suppressed relative to the elastic case. Nevertheless, for the $\mathcal{O}_{1}^{s}$ and $\mathcal{O}_{11}^{s}$ operators, which do not depend explicitly on the inelastic transverse velocity $\vec{v}_{\mathrm{inel}}^{\,\perp}$ defined in Eq.~(\ref{u_inelastic}), the signal-to-background ratio remains clearly distinguishable from the relative background uncertainty. This leads to improved sensitivity compared to LUX--ZEPLIN for these operators. In contrast, for the $\mathcal{O}_{5}^{s}$ and $\mathcal{O}_{8}^{s}$ operators, which depend explicitly on $\vec{v}_{\mathrm{inel}}^{\,\perp}$, the signal-to-background ratio is comparable to the background uncertainty, resulting in only a modest improvement over LUX--ZEPLIN. For a mass splitting of $\delta_{m}=100~\mathrm{keV}/c^{2}$, close to the upper range accessible to paleo-detectors, gypsum achieves sensitivity comparable to LUX--ZEPLIN for the $\mathcal{O}_{1}^{s}$ and $\mathcal{O}_{11}^{s}$ operators, while sensitivity to $\mathcal{O}_{5}^{s}$ and $\mathcal{O}_{8}^{s}$ is weaker due to the reduced signal-to-background ratio. Notably, paleo-detectors can probe lower DM masses than conventional experiments, setting constraints for $m_{\chi}\gtrsim30~\mathrm{GeV}/c^{2}$ when $\delta_{m}=50~\mathrm{keV}/c^{2}$ and for $m_{\chi}\gtrsim200~\mathrm{GeV}/c^{2}$ when $\delta_{m}=100~\mathrm{keV}/c^{2}$.

In this section, we have presented the projected 90\% confidence level exclusion limits on isoscalar WIMP--nucleon NREFT coupling constants for both elastic and inelastic scattering.
For elastic scattering, paleo-detectors with sufficiently high read-out resolution are projected to probe WIMP--nucleon couplings below current limits for $m_\chi \lesssim 10~\mathrm{GeV}/c^2$. At higher masses, $m_\chi \gtrsim 10~\mathrm{GeV}/c^2$, high-exposure scenarios enable constraints on NREFT couplings that are comparable to or stronger than those from conventional experiments, particularly for minerals such as gypsum.
For inelastic scattering, paleo-detectors are expected to improve sensitivity over conventional experiments for $m_\chi \gtrsim 50~\mathrm{GeV}/c^2$ and $\delta_m \lesssim 50~\mathrm{keV}/c^2$ for most NREFT operators. For larger mass splittings, $\delta_m \sim 100~\mathrm{keV}/c^2$, conventional experiments perform better due to heavier target nuclei, while paleo-detectors lose sensitivity for $\delta_m \gtrsim 100~\mathrm{keV}/c^2$.

\section{Conclusion} \label{conclusion}

In this contribution, we summarize the projected sensitivities of paleo-detectors—natural minerals that record nuclear damage tracks over geological timescales—to WIMP--nucleon interactions described within the framework of a non-relativistic effective field theory (NREFT), based on the results of Ref.~\cite{Theodosopoulos:2026ehn}. We considered both elastic and inelastic scattering with isoscalar couplings, and included the dominant backgrounds from solar, supernova, and atmospheric neutrinos, as well as radiogenic processes modeled following Ref.~\cite{Baum:2021jak}. Using a profile-likelihood approach, we derived projected $90\%$ confidence-level upper limits on the isoscalar NREFT coupling constants.

Our analysis focused on two representative readout scenarios: a high-resolution (HR) scenario with spatial resolution $\sigma_x = 1~\mathrm{nm}$ and sample mass $M = 10~\mathrm{mg}$, and a high-exposure (HE) scenario with $\sigma_x = 15~\mathrm{nm}$ and $M = 100~\mathrm{g}$. Results were presented for four representative minerals—gypsum, halite, olivine, and muscovite. Results for additional target materials, namely sinjarite, epsomite, phlogopite, and nchwaningite, can be found in the appendix of Ref.~\cite{Theodosopoulos:2026ehn}.

The projected paleo-detector sensitivities were compared with existing constraints from the XENON100~\cite{XENON:2017fdd}, LUX--ZEPLIN~\cite{LZ:2023lvz}, and PandaX--II~\cite{PandaX-II:2018woa} experiments, as well as with the $95\%$ Bayesian credible region reported by SuperCDMS~\cite{SuperCDMS:2022crd}. For DM masses in the range $1$--$10~\mathrm{GeV}/c^{2}$, where neutrino backgrounds dominate and recoil track lengths are short, paleo-detectors operating in the HR scenario exhibit superior sensitivity compared to conventional direct-detection experiments. For heavier DM masses, $m_{\chi}\sim10~\mathrm{GeV}/c^{2}$ up to several TeV, radiogenic backgrounds become dominant and recoil tracks are longer; in this regime, several paleo-detectors in the HE scenario achieve sensitivities comparable to or exceeding those of existing experiments for most NREFT operators. 

Throughout this analysis, we assumed a one-to-one correspondence between the observed track length $x_T$ and the nuclear recoil energy $E_R$ for DM masses in the range $1~\mathrm{GeV}/c^{2}$ to $5~\mathrm{TeV}/c^{2}$. As shown in Ref.~\cite{Fung:2025cub}, this approximation is accurate within this mass window. An important extension of this work is the exploration of lighter DM masses, for which this mapping becomes inaccurate. In that regime, a recoil with fixed energy can produce a distribution of track lengths, and the probability of track formation must be modeled explicitly following the approach of Ref.~\cite{Fung:2025cub}.

Finally, paleo-detectors offer an opportunity not only to constrain DM interactions, but also to discriminate between different NREFT operators. While operator discrimination in conventional direct-detection experiments often relies on recoil energy and directional information~\cite{Kavanagh:2015jma}, paleo-detectors could instead exploit their exceptionally large exposures and detailed track--length spectra to assess how well different operator hypotheses, or combinations thereof, describe a potential DM signal. This provides a promising avenue for future studies.

\acknowledgments

The author is grateful to the organizers of the Corfu2025 Workshop on the \textit{Standard Model and Beyond} for the opportunity to present this work. The author also acknowledges support from the University of Texas at Austin and thanks Katherine Freese, Chris Kelso, and Patrick Stengel for valuable discussions and collaboration.

\end{document}